\documentclass[%
 reprint,
 amsmath,amssymb,
 aps,
]{revtex4-2}

\makeatletter
\renewcommand*{\@fnsymbol}[1]{\ensuremath{\ifcase#1\or \dagger\or *\or \ddagger\or
   \mathsection\or \mathparagraph\or \|\or **\or \dagger\dagger
   \or \ddagger\ddagger \else\@ctrerr\fi}}
\makeatother

\usepackage{graphicx}
\usepackage{dcolumn}
\usepackage{bm}
\usepackage{hyperref}

\begin{document}

\title{Heterogeneous integration of solid state quantum systems with a foundry photonics platform}

\author{Hao-Cheng Weng}
 \altaffiliation{haocheng.weng@bristol.ac.uk}
\affiliation{Quantum Engineering Technology Labs, H. H. Wills Physics Laboratory and Department of Electrical and Electronic Engineering, University of Bristol, Bristol}

\author{Jorge Monroy-Ruz}
\affiliation{Quantum Engineering Technology Labs, H. H. Wills Physics Laboratory and Department of Electrical and Electronic Engineering, University of Bristol, Bristol}

\author{Jonathan C. F. Matthews}
\affiliation{Quantum Engineering Technology Labs, H. H. Wills Physics Laboratory and Department of Electrical and Electronic Engineering, University of Bristol, Bristol}

\author{John G. Rarity}
\affiliation{Quantum Engineering Technology Labs, H. H. Wills Physics Laboratory and Department of Electrical and Electronic Engineering, University of Bristol, Bristol}

\author{Krishna C. Balram}
\affiliation{Quantum Engineering Technology Labs, H. H. Wills Physics Laboratory and Department of Electrical and Electronic Engineering, University of Bristol, Bristol}

\author{Joe A. Smith}
 \altaffiliation{j.smith@bristol.ac.uk}
\affiliation{Quantum Engineering Technology Labs, H. H. Wills Physics Laboratory and Department of Electrical and Electronic Engineering, University of Bristol, Bristol}

\date{\today}

\begin{abstract}
Diamond colour centres are promising optically-addressable solid state spins  that can be matter-qubits, mediate deterministic interaction between photons and act as single photon emitters. Useful quantum computers will comprise millions of logical qubits. To become useful in constructing quantum computers, spin-photon interfaces must therefore become scalable and be compatible with mass-manufacturable photonics and electronics. Here we demonstrate heterogeneous integration of NV centres in nanodiamond with low-fluorescence silicon nitride photonics from a standard 180 nm CMOS foundry process. Nanodiamonds are positioned over pre-defined sites in a regular array on a waveguide, in a single post-processing step. Using an array of optical fibres, we excite NV centres selectively from an array of six integrated nanodiamond sites, and collect the photoluminescence (PL) in each case into waveguide circuitry on-chip. We verify single photon emission by an on-chip Hanbury Brown and Twiss cross-correlation measurement, which is a key characterisation experiment otherwise typically performed routinely with discrete optics. Our work opens up a simple and effective route to simultaneously address large arrays of individual optically-active spins at scale, without requiring discrete bulk optical setups. This is enabled by the heterogeneous integration of NV centre nanodiamonds with CMOS photonics.
\end{abstract}

\maketitle

Solid state atom-like systems such as the nitrogen-vacancy (NV) centre in diamond show great promise for quantum information \cite{pezzagna2021quantum,ruf2021quantum}. Entanglement state generation \cite{bernien2013heralded} and distillation \cite{kalb2017entanglement} have been realised in spin-interfaced quantum networks. Optically-addressable spin states operate as quantum registers and memories for quantum computing \cite{waldherr2014quantum} and communications \cite{pompili2021realization}. In addition, NV centres have exceptional characteristics for quantum sensing with high resolution and high signal-to-noise ratio imaging of electromagnetic fields \cite{gross2017real, dolde2011electric} where introducing spins into photonic waveguides is a prerequisite for emerging applications in quantum sensing \cite{hoese2021integrated,bai2020fluorescent}. 
Bringing solid state systems into foundry photonics 
would provide the missing component in the integrated quantum photonics toolkit \cite{moody20222022}. 

Combining atom-like systems into an integrated quantum processor remains a major technological hurdle to date, requiring precise photonic/spin manipulations at scale, and the creation of centres at nanoscale resolution \cite{awschalom2018quantum}. Miniaturisation of photonics and electronics have enabled selective control and manipulation on the same chip \cite{kim2019cmos,zhang2017selective}. A pick-and-place method has been adopted to combine diamond hosted colour centres with integrated photonics \cite{wan2020large, chakravarthi2022hybrid}. However, challenges remain to achieve integration using foundry-manufacturable photonics in a scalable process. In particular, the need to identify \cite{narun2022efficient, kudyshev2020rapid} and manipulate stochastic emitters \cite{wolters2010enhancement, van2009nanopositioning} requires measurement over sample areas (${\approx}\text{mm}^2$) with nanoscale precision and is a highly sensitive and time-intensive technique. 

\begin{figure*}[ht]
\centering
\includegraphics[width=\textwidth]{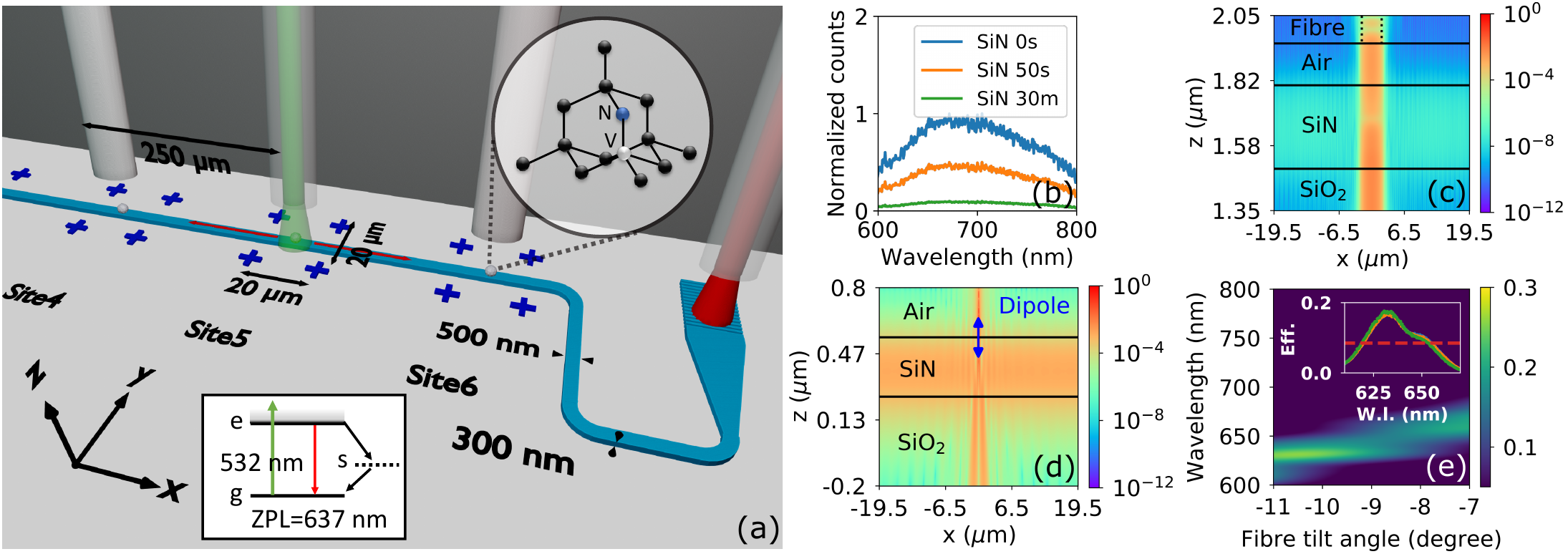}
\caption{\textbf{Integration scheme and photonic design.} (a) Photonic integration platform of NV centres in NDs. Diagram not to scale. NDs (spheres in white) are positioned over six separate sites on top of a silicon nitride waveguide (in light blue). Each site is defined by four alignment markers (in dark blue). With an array of fibres coupled to the chip (figure upper part), NV centres are excited by 532 nm (green) laser by selecting one of the middle fibres. Radiated decay of NV is accompanied by PL (in red) emission that is coupled to the waveguide and collected by fibres at the two ends through grating couplers. The NV lattice structure and energy levels are shown in the inset. The coordinate system indicated in the figure will be used throughout the article. (b) Photobleaching SiN fluorescence. The unbleached/0-second, 50-second, and 30-minute bleaching result is seen via the SiN fluorescence spectrum scaled according to the bleaching process (Section C in the supplementary). Measurement is performed using confocal microscopy. (c) Modelling of spatial filtering of the pump field. 532 nm laser output from the fibre core (in dashed line), shines directly on the bare SiN waveguide with SiO$_2$ substrate. Power of electromagnetic field is recorded and also note the highly expanded z-scale. (d) Coupling of a NV centre to the waveguide mode. NV centre is modeled by an electric dipole oriented in the z direction at the centre of waveguide. The dipole emission is coupled (from the top) to the SiN waveguide. Power of electromagnetic field is shown in this figure. (e) Efficiency of single grating coupler by design. Inset shows the experimental results for three devices on the same chip measured at 9 degrees fibre tilt.}
\label{fig:scheme}
\end{figure*}

Our approach is to combine NV centres in nanodiamonds (nanoscale inclusions of diamond or NDs) with a standard foundry process. With this heterogeneous approach \cite{moody20222022}, we combine the advantages of diamond as an emitter host with the maturity of a foundry photonics platform. To successfully integrate NV centres with an existing process, it is necessary to select a photonics platform that exhibits moderate refractive index contrast, visibility at the wavelength of interest, low propagation loss, with mature characteristics and devices. For this, we use IMEC's BioPIX silicon nitride (SiN) platform with a refractive index of $n = 1.89$ and a propagation loss of 0.9 dB/cm at 638 nm \cite{masood2020pix4life}. Of particular importance, this process uses nitrogen-rich SiN critical for its low autofluorescence and compatibility with NV centres \cite{smith2020single}.

\section{Results}

Here, the NV centre integration involves a single post-processing step with NDs precisely ($\pm 200$ nm w.r.t. the waveguide centre) and deterministically positioned over millimetre scales, with respect to foundry-defined markers in the SiN (Fig.\ref{fig:scheme} (a)). NV centres in NDs are chosen for the advantage of precise positioning \cite{van2009nanopositioning, schell2011scanning, ampem2009nano} and moderate quantum coherence. 

The manufactured chip consists of four identical devices, each being an air-cladded SiN strip waveguide of 500 nm width and 300 nm thickness on top of a silicon dioxide substrate. The waveguides are bent 90 degree at the ends and tapered to mode match the grating couplers. Grating couplers are designed to match PM630 fibres (4 \textmu m mode field diameter). Using nitrogen-rich SiN prevents the pump from creating excess background fluorescence in the waveguide, which greatly diminishes the signal-to-noise ratio in stoichiometric SiN \cite{smith2020single}. In addition, under strong 532 nm excitation, we observe the SiN can be bleached permanently, resulting in even lower autofluorescence between 600 nm and 800 nm. We compare the autofluorescence during different stages of the bleaching process (Fig.\ref{fig:scheme} (b)). At 2 mW pump power, a 50-second bleaching reduces 50\% of the autofluorescence and a 90\% reduction is observed after 30 minutes.

Six excitation sites are evenly distributed along the waveguide with 250 \textmu m separation. This enables a scheme to excite NV centres and collect photoluminescence (PL) using a v-groove array of eight fibres (OZ Optics) that couples to the chip. NV centres are excited off-resonantly by a 532 nm pump laser through one of the six excitation fibre channels to emit single photons from lithographically-defined locations. Through this all-fibre excitation and collection scheme, we eliminate the need for vibration-sensitive bulk optics. The small cross section between the 4 \textmu m excitation beam and the 500 nm-wide waveguide, as well as the orthogonal pump-probe design, result in a strong spatial filtering of the pump. With the bare waveguide, coupling of 532 nm laser to the waveguide and thus fluorescence of SiN in the channel are strongly suppressed by -70 dB as shown in Fig.\ref{fig:scheme} (c) (detailed in Section C of the supplementary).

Relaxation from the excited NV centre is accompanied by single photon emission in the zero-phonon line (637 nm) and phonon sideband which is coupled to the evanescent field of the SiN waveguide, and guided along its cross-section (Fig.\ref{fig:scheme} (d)). The moderate refractive index contrast of SiN-on-silica results in good coupling between the NV centre in NDs and the waveguide channel. We find an average of 9\% (and a maximum of 28\%) coupling efficiency over different positions and orientations of NV centres. Our method improves on integration in low-contrast platforms such as silica \cite{fujiwara2017fiber} or laser-written diamond waveguides \cite{hadden2018integrated}. PL is guided and collected by fibres at the two ends through grating couplers. The grating coupler is designed around the zero-phonon line with a maximum collection efficiency of 0.25 at 630 nm (Fig.\ref{fig:scheme} (e)). A comparable experimental result shows the maximum efficiency of 0.17 (-7.7 dB) at 630 nm when the fibre is tilted at 9 degrees. The coupling efficiency is repeatable across all measured devices. Inset of Fig.\ref{fig:scheme} (e)  shows result of three devices highly overlapped. The grating couplers exhibit a 40 nm bandwidth (FWHM shown by red dashed line), which helps to filter the 532 nm pump and other background noise.

\begin{figure*}[htb!]
    \centering\includegraphics[width=\textwidth]{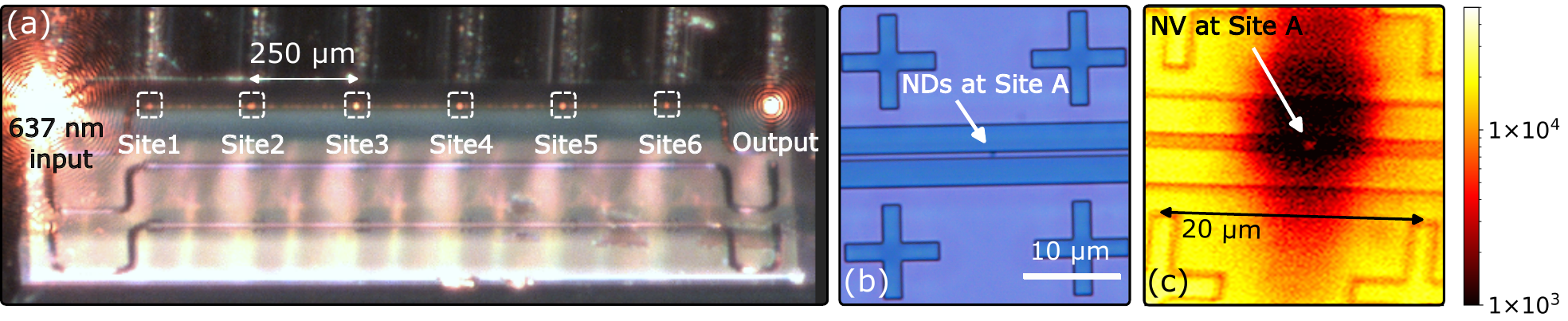}
    \caption{\textbf{Nanodiamond (ND) positioning on a foundry photonics chip.} (a) NDs observed over six sites (in white boxes) of a single device on chip. A fibre array (figure upper part) is coupled to the photonic chip (figure lower part). NDs on the waveguide are observed by scattered 637 nm laser. (b) An optical microscope image of Site A where NDs are positioned centrally on top of the waveguide. (c) Confocal microscope PL image of Site A. A contrast between NV PL (3000 counts/s) and the bleached SiN ($<$ 400 counts/s) locates the position of NDs on the waveguide. }
    \label{fig:deposition}
\end{figure*}

 Integration of NV centres in NDs is achieved in a single lithography deposition step. Each alignment site is defined by four crosses in the SiN layer, in the centre of a 20 \textmu m by 20 \textmu m square. A polymethyl methacrylate (PMMA) resist mask is first spincoated on the photonic chip. A 4 \textmu m long and 400 nm wide region (x$\times$y) is patterned in the middle of each alignment site. The 400 nm width limits NDs to the centre of waveguide while the long 4 \textmu m region covers the full excitation region of the 532 nm laser. We use electron beam lithography (Raith Voyager) for PMMA patterning although a similar technique with a photoresist would be well suited for standard photolithography. After resist development, NDs in solution are deposited on the chip and left to dry to allow the solvent to evaporate. The PMMA mask is then removed and NDs remain only over the defined regions on top of the waveguides.  We use NDs of around 50 nm diameter, milled from high pressure high temperature diamond (Nabond). The concentration of nitrogen atoms shows an averaged probability of 2\% to find a single NV centre in a single ND \cite{knowles2014observing}. 

An exemplar device is shown in Fig.\ref{fig:deposition} (a). The positioned NDs are identified by scattered light when 637 nm laser is coupled in through the left grating coupler and out through the right one. The high yield results in NDs at each site (labelled Site1 to Site6). For a site (Site A) studied in experiment, the positioned NDs are inspected under optical microscope (Fig.\ref{fig:deposition} (b)) and found well-aligned on the top of the waveguide at the centre of the site. This is also supported by a fine confocal-scanned image (Fig.\ref{fig:deposition} (c)) with 1 \textmu m excitation spot size. For a chip of 24 positions over 4 devices, our method presents a 70\% yield rate of ND positioning either by identifying through the scattered 637 nm laser or direct observation under optical microscope. It should be stressed that near unity yield rate of ND positioning is possible \cite{heffernan2017nanodiamond}. However, assuming a constant filling rate, our deposition area is chosen to maximise the probability distribution of finding one NV centre per nanodiamond site.

\begin{figure*}[ht]
    \centering\includegraphics[width=15cm]{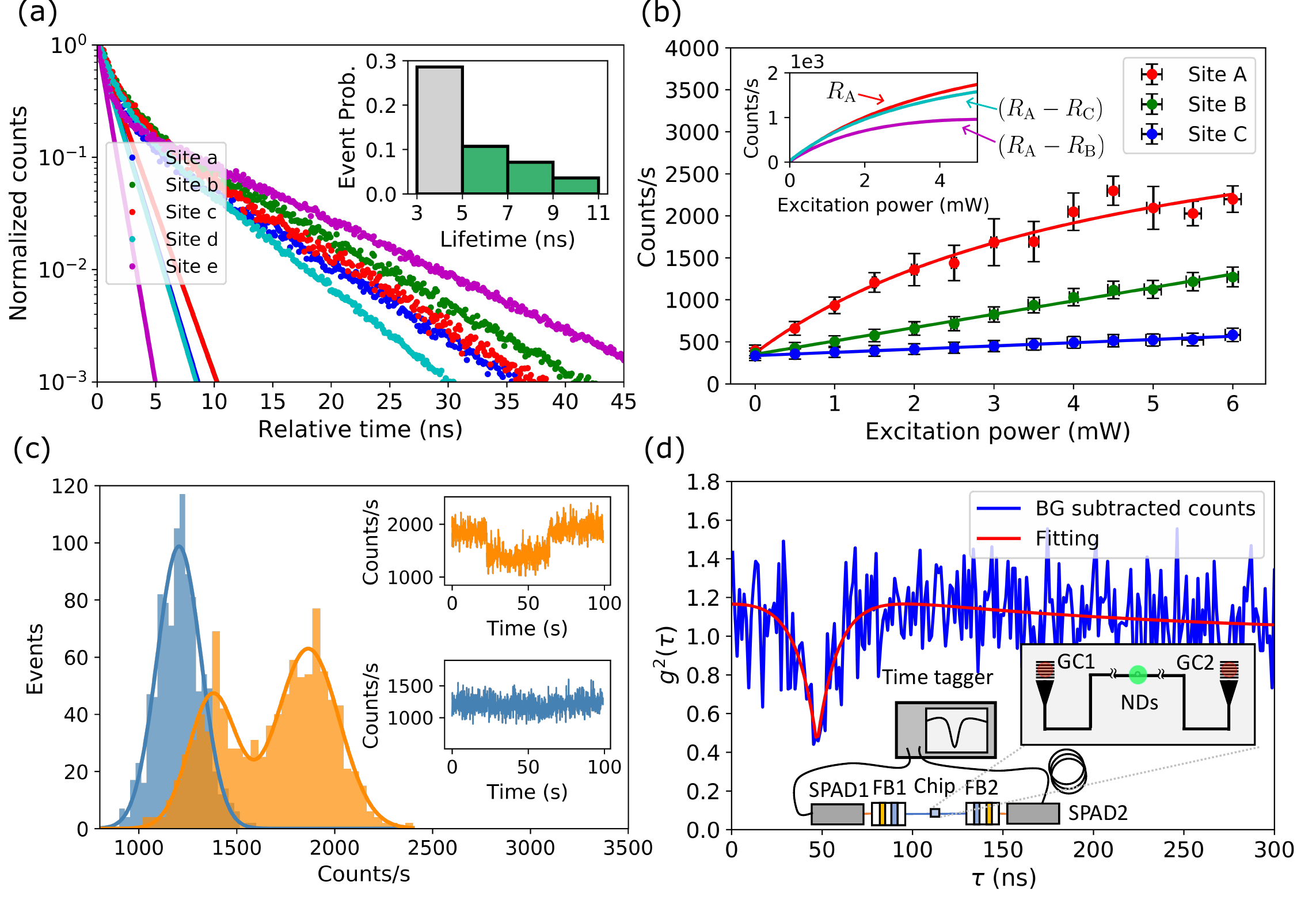}
    \caption{\textbf{Coupled device results.} (a) PL lifetime of five sites containing NV centres. Data points with detector dark count subtracted. Solid lines are fitted fast-decay parts. Inset: statistics of lifetimes for all sites containing NDs. (b) Saturated emission. Error bars present standard deviation of detected counts per second and excitation power. The data points are fitted to Eq.\ref{Eq_sat}. Inset: Saturated emission with background fluorescence subtracted. (c) Blinking in the PL. The count rate per second is recorded for 100s to show a histogram of count rates and the PL time traces. The blinking (histogram in orange, measured at 3 mW pump power) and non-blinking (histogram in blue, measured at 1.5 mW pump power) statistics are compared. (d) HBT experiment. The data is background subtracted and fitted with Eq.\ref{Eq_anti}. Inset is the schematic HBT measurement setup. Single photon emission is collected by two grating couplers (GC) separately and fibre-coupled to filter boxes (FB) containing a notch filter and long-pass filter. Single photon counting is time correlated by the time tagger.}
    \label{fig:results}
\end{figure*}

To identify that the site-positioned ND contains a NV centre that couples to the device, and that this device is successful in both exciting and sufficiently filtering the pump, time-domain photodynamics are measured through the grating output. PL from NV centres exhibits a characteristically long-lived lifetime of tens of nanoseconds during relaxation from the excited state to ground state \cite{berthel2015photophysics,smith2020single}. We excite the candidate site by a 532 nm pulsed laser through its input port and the lifetime is measured by coincidence counting between the trigger of the laser and counts detected on a single photon avalanche diode (SPAD) fibre coupled to the grating output. The time resolved PL is fitted to a bi-exponential function
\begin{equation}
    I(t-t_0)=I_1 \ e^{-(t-t_0)/t_1}+I_2 \ e^{-(t-t_0)/t_2}+I_{\textrm{bias}},
    \label{Eq_saturation}
\end{equation}
describing two exponential decays and a constant bias. The decay terms model the NV PL and the background signal of different  lifetimes ($t_1$ and $t_2$) while the constant bias is a result of detector dark counts. $I_1$, $I_2$, and $I_{\textrm{bias}}$ are the relative fractions and the fitting is a function of relative time $t-t_0$ where $t_0$ is the pulsed trigger. 

 For five different sites, the lifetimes (data points) are compared with their fitted fast decaying part (solid lines), as shown in Fig.\ref{fig:results} (a). The contribution of $I_{\textrm{bias}}$ as a result of detector dark counts has been subtracted. The fitted emission is dominated by a slow decay which is evidence of PL from NV centres, with lifetime ranging from $5-11$ ns, as compared with the fast decay term. This lifetime is evidence of good coupling between the NV centre and the waveguide \cite{smith2020single,smith2021nitrogen}. The fast decay contribution is a result of background fluorescence, mainly from the excited fibre \cite{schroder2011ultrabright,schroder2012nanodiamond, patel2016efficient,duan2019efficient} plus minimal fluorescence contribution from SiN (details in supplementary Section C). The background fluorescence falls in the range of $1-5$ ns. The statistics of PL lifetime from NV centres are plotted in the Fig.\ref{fig:results} (a) inset with green bars. PL lifetimes measured shorter than 5 ns are coloured in grey as potentially mixed with the background fluorescence term. Conditional on identifying the slow decay PL in a lifetime measurement (44\% of the total events fall within green bars in the inset figure), we report on average a 30\% ($0.44 \times 0.70$) yield rate of at least one NV centre per excitation site.

Another characteristic of NV centres is saturation under strong pumping \cite{kurtsiefer2000stable,schroder2011ultrabright}.  We study saturation by exciting the site with increasing power and detecting PL through one of the grating outputs. In Fig.\ref{fig:results} (b), data is fitted to
\begin{equation}
    R(P)=R_{\textrm{sat}}\  \frac{P}{(P_{\textrm{sat}}+P)}+aP+b,
    \label{Eq_sat}
\end{equation}
where $R$ describes the emission rate as a function of the laser power $P$ saturating at $R_{\textrm{sat}}$ with $P_{\textrm{sat}}$. Here $a$ models linear response to the  power from the background and $b$ constant dark counts. Three sites on the same device are measured to give reference background fluorescence contributions. NDs are observed optically and by scattered 637 nm laser for both Site A and Site B. A slow PL (of 10 ns lifetime) is measured at Site A but not Site B. The results are compared with a bare waveguide region without NDs (Site C). From the fit, Site A ($R_\textrm{A}$) clearly saturates, giving further evidence that the signal is dominated by NV centre PL. Counts at Site B ($R_\textrm{B}$) and Site C ($R_\textrm{C}$) show linear responses as expected for background fluorescence. The background is much higher with the presence of NDs on top of the waveguide, assumed due to additional scatter into the waveguide channel (see supplementary Section C). Considering the ND presence, $R_\textrm{C}$ is regarded as a lower bound to the noise level and a fair noise estimation at Site A should be $R_\textrm{B}$. In Fig.\ref{fig:results} (b) inset, the red curve shows $R_\textrm{A}$ in comparison with the cyan curve ($R_\textrm{A}-R_\textrm{C}$) and the magenta curve ($R_\textrm{A}-R_\textrm{B}$). The curves are fitted by the saturation model $R(P)= R_{\textrm{sat}} P/(P_{\textrm{sat}}+P)$ alone. The net saturated emission is then recovered (combining both grating outputs) with $R_{\textrm{sat}}=2600-5500$ counts/s and $P_{\textrm{sat}}=1.8-3.7$ mW. The intervals are lower bounded by fitting $(R_\textrm{A}-R_\textrm{B})$ and upper bounded by $(R_\textrm{A}-R_\textrm{C})$. This saturated power corresponds to $1.4\times 10^8-3.0\times 10^8 \ \textrm{W/cm}^2$ power density.

In addition to lifetime and saturation, NV$^-$ in small sized NDs has been shown to be exhibit intermittency in the PL (blinking) under strong excitation of 532 nm. This can be due to the transition between NV$^-$ and NV$^0$ or other non-radiative decay routes \cite{inam2013tracking, bradac2010observation}. Under high excitation powers, we measure blinking at Site A through the grating coupler, contributing to further evidence of NV centre dominated emission. NV centres are excited by CW 532 nm laser while the detected count rate per second is recorded for a 100-second interval. In Fig.\ref{fig:results} (c), the count-rate histogram shows double peaks (in orange) when the NV centres are excited with 3 mW pump power. This is compared with a single peak (in blue) under 1.5 mW pump. The blinking can also be seen by the time trace of detected count rates. A sudden decrease in the brightness is followed by recovery after several tens of seconds (Fig.\ref{fig:results} (c) insets).

The de-facto method of determining whether emission in a channel arises from a \emph{single} NV centre is by observing anti-bunched statistics through a HBT experiment, where the time-resolved cross-correlation of detection events from two output ports is recorded and normalized to calculate the second-order autocorrelation function $g^2(\tau)$. Here, $\tau$ is the relative delay between the two detection events after splitting and a dip is expected at $g^2(\tau=0)$. The value at $g^2(0)$ scales for multi-emitters following $g^2(0)=1-1/n$, where n is the number of NV centres \cite{kurtsiefer2000stable, berthel2015photophysics}. As a result, the visibility is diminished when multiple NV centres contribute to the emission. With this device, we perform an on-chip HBT experiment by which a beamsplitter is formed from the NV centre coupling to the separate paths of the waveguide ($-x$ and $x$). These paths are  collected by grating couplers at either end of the device (inset of Fig.\ref{fig:results} (d)). One detector is delayed, shifting the dip from $\tau=0$ to $\tau=\tau_0$ for the resolving of the whole dip. In Fig.\ref{fig:results} (d), the on-chip HBT for Site A is presented. The dip at $\tau_0=48$ ns evidences the quantum statistics of single photon emission. The $g^2(\tau)$ is calculated with background estimated and subtracted. Considering the signal-to-noise ratio $\sigma=S/(S+N)$, where $S$ and $N$ is respectively the signal and noise, the autocorrelation function under background noise is related to that without background by $g_{\textrm{gb}}^2(\tau)=1-\sigma^2+\sigma^2 g^2(\tau)$ \cite{Brouri2000Photon}. From the saturation measurement in Fig.\ref{fig:results} (b), we estimate $\sigma=(R_\textrm{A}-R_\textrm{B})/R_\textrm{A}=0.42$ at 6 mW pump. We are able to determine that 95\% of this noise originates from the fluorescence generated in the excitation fibre (see supplementary Section C), with a residual amount contributed by the pump in the waveguide. The result is fitted to a three-level-system model \cite{patel2016efficient,berthel2015photophysics},
\begin{equation}
    g^2(\tau-\tau_0)=1-p_{\textrm{ge}} \ e^{-(\tau-\tau_0)/\tau_{\textrm{ge}}}+p_{\textrm{s}} \ e^{-(\tau-\tau_0)/\tau_{\textrm{s}}},
    \label{Eq_anti}
\end{equation}
where the first term describes transition between the excited state and ground state while the second term with a shelving state. Here, $\tau_{\textrm{ge}}$ ($\tau_{\textrm{s}}$) and $p_{\textrm{ge}}$ ($p_{\textrm{s}}$) are the corresponding lifetime and intensity for each term. From the fit, $\tau_{\textrm{ge}}$ is found to be 11 ns which corresponds well with the 10 ns lifetime measured at Site A. The $\tau_{\textrm{s}}$ is found to be 186 ns in good agreement with the literature \cite{berthel2015photophysics}. With the background well-estimated and corrected, we measure $g^2(0)=0.48$. The same site is measured with a confocal setup showing $g^2(0)=0.46$ after background subtraction (supplementary Section B), in good agreement. A $g^2(0)\approx 0.5$ implies that two NV centres are excited at this site.

\begin{figure}[ht]
    \centering\includegraphics[width=\linewidth]{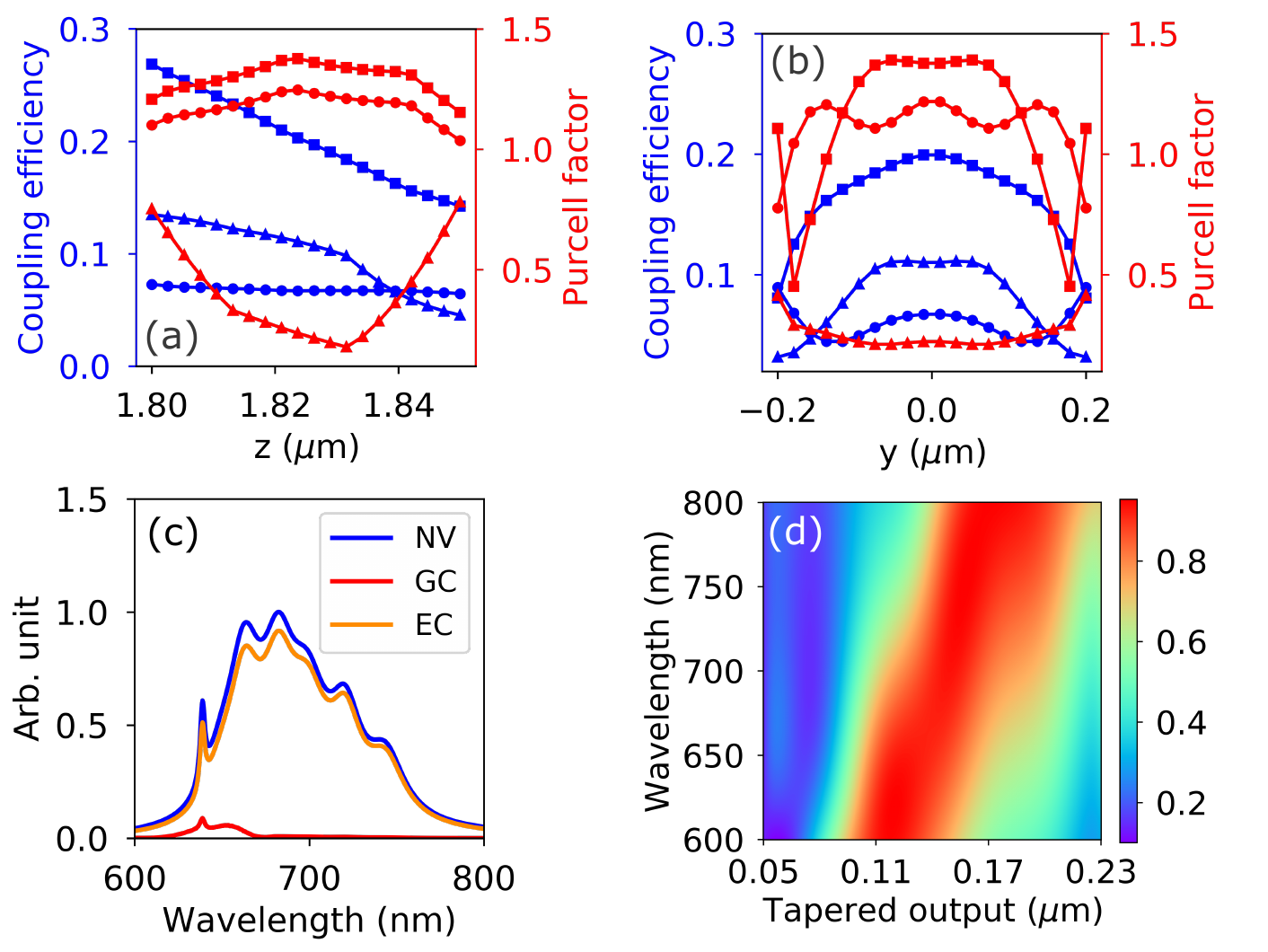}
    \caption{\textbf{Coupling efficiency and signal-to-noise ratio improvement.} (a-b) Coupling efficiency from NV centre to the waveguide and the Purcell factor. NV centre is modeled by an electric dipole source embedded in a 4 \textmu m $\times$ 400 nm $\times$ 50 nm (x$\times$y$\times$z) region of refractive index 2.4 (diamond) to simulate the nanodiamond deposition area. The coupling is studied when NV is at different depth (in the z direction) and displaced from the centre (in the y direction). The dipole is oriented in the x, y, and z direction for square, round, and triangle markers, respectively. (c) Transmission spectrum of grating couplers (GC) and edge couplers (EC) as compared to the NV centre PL spectrum. (d) Mode-overlap \cite{snyder1983optical} between the waveguide and fibre mode. The overlapping is calculated for different wavelength and waveguide tapered output.}
    \label{fig: sim_n_imp}
\end{figure}

\section{Discussion}
In this first device, the photon flux measured through the waveguide channel is low compared to a high-NA confocal microscope, resulting in long integration times for the two-channel HBT measurement. To improve this, the coupled device should be properly modelled. We can describe the detected emission rate $R_{\textrm{det}}$ of Site A by:
\begin{equation}
    R_{\textrm{det}}=\eta_{\textrm{q}}    \ (1/\tau_{\textrm{rad}}) \ \eta_{\textrm{wg}} \ \eta_{\textrm{grating}} \ \eta_{\textrm{det}}
    \label{eq_brightness}
\end{equation}
Here, the internal quantum efficiency of the NV centre $\eta_{\textrm{q}}$ captures decay through non-radiative routes (including the observed blinking). In sub-100nm NDs, owing to various surface processes, this can range from 0.02 to 0.25 \cite{mohtashami2013suitability,plakhotnik2018nv} so we take a value of $\eta_{\textrm{q}}=\sqrt{0.02\times 0.25}=0.07$. The radiative lifetime $\tau_{\textrm{rad}}$ of the NV centre is strongly dependent on the surrounding geometry as the local density of optical state (LDOS) can be modified and the emission Purcell enhanced or suppressed \cite{inam2013emission}. We can quantify this by the Purcell factor $F=P_{\textrm{act}}/P_{\textrm{d}}$, from the actual power radiated normalised by the power radiated in bulk diamond. Simulations of the device finds $F$ ranges from 0.2 to 1.4 depending on position and dipole orientation of the NV centre within the deposition region (Fig.\ref{fig: sim_n_imp} (a) and Fig.\ref{fig: sim_n_imp} (b)). This variation supports the range measured in Fig.\ref{fig:results} (a) with $\tau_{\textrm{rad}}$  measured as 10 ns for Site A. Coupling to the waveguide $\eta_{\textrm{wg}}$ also depends on the orientation of NV centre and its position. Modeled in Fig.\ref{fig: sim_n_imp} (a) and Fig.\ref{fig: sim_n_imp} (b), $\eta_{\textrm{wg}}$ ranges from $0.03-0.27$ collecting from both waveguide ends with an averaged value $\eta_{\textrm{wg}}=\sqrt{0.03\times0.27}=0.09$. 
External efficiencies include the grating couplers designed to capture the zero-phonon line. A reference PL spectrum of the NV centre \cite{albrecht2013coupling} is shown with its broad phonon sideband in Fig.\ref{fig: sim_n_imp} (c), in comparison with the grating transmission. The area under the curve gives a total efficiency $\ \eta_{\textrm{grating}}=0.02$ over the spectrum. The final term $\eta_{\textrm{det}}=0.21$ combines the off-chip efficiency, including the measured transmission through the filters and the detector efficiency. From this model, $R_{\textrm{det}}$ of a single NV centre at Site A is estimated as 2650 counts/s. This falls in the range of measured saturated counts for a single NV centre of 1300-2750 counts/s, with the assumption that Site A contains two emitters.
This is compared with 40-160 counts/s$\cdot$mW background fluorescence and 350 counts/s detector dark counts estimated from Fig.\ref{fig:results} (b) at each grating coupler. 

Considering the contributions to the signal loss, the detected emission could be improved by several strategies. One is to pursue edge couplers as opposed to the grating coupler. An advantage of this moderate refractive index platform is that it can be well-matched to PM630 fibre without lenses or complicated structures. We design a process-compatible edge coupler by tapering the waveguide width from 500 nm to 150 nm (minimum feature size) and observe a good mode overlap (84\%) with the fibre (4 \textmu m diameter) at 637 nm. High efficiency ($>$75\%) is maintained over a broad bandwidth (Fig.\ref{fig: sim_n_imp} (d)). Compared to grating couplers, this leads to a 35 fold improvement in flux across the NV spectrum (Fig.\ref{fig: sim_n_imp} (c)). To improve the waveguide mode overlap, we could overcoat the ND in silicon nitride \cite{smith2020single}. We can also under etch the substrate to reduce leakage \cite{fehler2019efficient}. In addition, we could pursue slot waveguides \cite{Hiscocks2009Slot} or nanophotonic cavities to increase Purcell enhancement \cite{wolters2012coupling, van2011deterministic}. 

Towards quantum photonic applications, we must isolate the remaining background from the measured signal. This contribution is not limited by the low-noise silicon nitride waveguide with fluorescence in the excitation fibre corresponding to 95\% of the noise (supplementary Section C). By shortening the fibre, adding laser line filters at 532 nm, and switching to photonic crystal fibres \cite{fujii2011fusion}, we can expect strong attenuation of this fluorescence. Shown in Fig.\ref{fig:results} (b), the background observed is proportional to the excitation power. In our scheme, the weak overlap with the 4 \textmu m mode field diameter excitation fibre and the NV centre requires significant additional power. In future, moving to a lens/tapered fibre with a mode field diameter of 1 \textmu m would result in a 16-fold suppression of the background fluorescence. Additionally, we estimate that the fluorescence coupled to the waveguide is tripled on the presence of the positioned NDs, leading to higher noise level observed in Fig.\ref{fig:results} (b). Considering different diamond geometries, including bulk diamond, could reduce scattering, as well as overgrowing index-matching silicon nitride \cite{smith2020single}.

We have demonstrated the integration of NV centres with foundry photonics where scalable and precise ND positioning is realised in a single post-processing step. We evidenced NV centre coupling through four experimental signatures. Although our work focuses on integrating NV centres, the idea can be generally adopted by other emitters \cite{lee2020integrated}, opening a route to chip-based integration. The all-integrated scheme provides mechanical stability compared to bulk optics or tapered fibres. This could also lead to stabler cryogenic operation, where difficulty of cooling ultrathin fibre tapers has been reported \cite{ji2023ultra,fujiwara2017fiber}. In addition, integrating multiple emitters on the same device demonstrates the idea of routing \cite{lee2022quantum} and multiplexing \cite{koong2020multiplexed} for quantum memories or quantum repeaters. Through foundry compatibility, a scalable platform is envisioned, enriched by the wide range of SiN photonics components and integrated electronics \cite{sharma2020review}.
\vspace{1em}

\section*{Acknowledgments}
 We thank A. Laing and L. Kling for useful suggestions. The IMEC BioPix 300 silicon nitride foundry run was funded by the HiSilicon Sponsorship Programme. This work was supported by EPSRC QC:SCALE EP/W006685/1. Post-processing was performed using equipment acquired through EPSRC QuPIC EP/N015126/1. 

\bibliography{References}

\begin{thebibliography}{53}%
\makeatletter
\providecommand \@ifxundefined [1]{%
 \@ifx{#1\undefined}
}%
\providecommand \@ifnum [1]{%
 \ifnum #1\expandafter \@firstoftwo
 \else \expandafter \@secondoftwo
 \fi
}%
\providecommand \@ifx [1]{%
 \ifx #1\expandafter \@firstoftwo
 \else \expandafter \@secondoftwo
 \fi
}%
\providecommand \natexlab [1]{#1}%
\providecommand \enquote  [1]{``#1''}%
\providecommand \bibnamefont  [1]{#1}%
\providecommand \bibfnamefont [1]{#1}%
\providecommand \citenamefont [1]{#1}%
\providecommand \href@noop [0]{\@secondoftwo}%
\providecommand \href [0]{\begingroup \@sanitize@url \@href}%
\providecommand \@href[1]{\@@startlink{#1}\@@href}%
\providecommand \@@href[1]{\endgroup#1\@@endlink}%
\providecommand \@sanitize@url [0]{\catcode `\\12\catcode `\$12\catcode
  `\&12\catcode `\#12\catcode `\^12\catcode `\_12\catcode `\%12\relax}%
\providecommand \@@startlink[1]{}%
\providecommand \@@endlink[0]{}%
\providecommand \url  [0]{\begingroup\@sanitize@url \@url }%
\providecommand \@url [1]{\endgroup\@href {#1}{\urlprefix }}%
\providecommand \urlprefix  [0]{URL }%
\providecommand \Eprint [0]{\href }%
\providecommand \doibase [0]{https://doi.org/}%
\providecommand \selectlanguage [0]{\@gobble}%
\providecommand \bibinfo  [0]{\@secondoftwo}%
\providecommand \bibfield  [0]{\@secondoftwo}%
\providecommand \translation [1]{[#1]}%
\providecommand \BibitemOpen [0]{}%
\providecommand \bibitemStop [0]{}%
\providecommand \bibitemNoStop [0]{.\EOS\space}%
\providecommand \EOS [0]{\spacefactor3000\relax}%
\providecommand \BibitemShut  [1]{\csname bibitem#1\endcsname}%
\let\auto@bib@innerbib\@empty
\bibitem [{\citenamefont {Pezzagna}\ and\ \citenamefont
  {Meijer}(2021)}]{pezzagna2021quantum}%
  \BibitemOpen
  \bibfield  {author} {\bibinfo {author} {\bibfnamefont {S.}~\bibnamefont
  {Pezzagna}}\ and\ \bibinfo {author} {\bibfnamefont {J.}~\bibnamefont
  {Meijer}},\ }\bibfield  {title} {\bibinfo {title} {Quantum computer based on
  color centers in diamond},\ }\href@noop {} {\bibfield  {journal} {\bibinfo
  {journal} {Applied Physics Reviews}\ }\textbf {\bibinfo {volume} {8}},\
  \bibinfo {pages} {011308} (\bibinfo {year} {2021})}\BibitemShut {NoStop}%
\bibitem [{\citenamefont {Ruf}\ \emph {et~al.}(2021)\citenamefont {Ruf},
  \citenamefont {Wan}, \citenamefont {Choi}, \citenamefont {Englund},\ and\
  \citenamefont {Hanson}}]{ruf2021quantum}%
  \BibitemOpen
  \bibfield  {author} {\bibinfo {author} {\bibfnamefont {M.}~\bibnamefont
  {Ruf}}, \bibinfo {author} {\bibfnamefont {N.~H.}\ \bibnamefont {Wan}},
  \bibinfo {author} {\bibfnamefont {H.}~\bibnamefont {Choi}}, \bibinfo {author}
  {\bibfnamefont {D.}~\bibnamefont {Englund}},\ and\ \bibinfo {author}
  {\bibfnamefont {R.}~\bibnamefont {Hanson}},\ }\bibfield  {title} {\bibinfo
  {title} {Quantum networks based on color centers in diamond},\ }\href@noop {}
  {\bibfield  {journal} {\bibinfo  {journal} {Journal of Applied Physics}\
  }\textbf {\bibinfo {volume} {130}},\ \bibinfo {pages} {070901} (\bibinfo
  {year} {2021})}\BibitemShut {NoStop}%
\bibitem [{\citenamefont {Bernien}\ \emph {et~al.}(2013)\citenamefont
  {Bernien}, \citenamefont {Hensen}, \citenamefont {Pfaff}, \citenamefont
  {Koolstra}, \citenamefont {Blok}, \citenamefont {Robledo}, \citenamefont
  {Taminiau}, \citenamefont {Markham}, \citenamefont {Twitchen}, \citenamefont
  {Childress} \emph {et~al.}}]{bernien2013heralded}%
  \BibitemOpen
  \bibfield  {author} {\bibinfo {author} {\bibfnamefont {H.}~\bibnamefont
  {Bernien}}, \bibinfo {author} {\bibfnamefont {B.}~\bibnamefont {Hensen}},
  \bibinfo {author} {\bibfnamefont {W.}~\bibnamefont {Pfaff}}, \bibinfo
  {author} {\bibfnamefont {G.}~\bibnamefont {Koolstra}}, \bibinfo {author}
  {\bibfnamefont {M.~S.}\ \bibnamefont {Blok}}, \bibinfo {author}
  {\bibfnamefont {L.}~\bibnamefont {Robledo}}, \bibinfo {author} {\bibfnamefont
  {T.~H.}\ \bibnamefont {Taminiau}}, \bibinfo {author} {\bibfnamefont
  {M.}~\bibnamefont {Markham}}, \bibinfo {author} {\bibfnamefont {D.~J.}\
  \bibnamefont {Twitchen}}, \bibinfo {author} {\bibfnamefont {L.}~\bibnamefont
  {Childress}}, \emph {et~al.},\ }\bibfield  {title} {\bibinfo {title}
  {Heralded entanglement between solid-state qubits separated by three
  metres},\ }\href@noop {} {\bibfield  {journal} {\bibinfo  {journal} {Nature}\
  }\textbf {\bibinfo {volume} {497}},\ \bibinfo {pages} {86} (\bibinfo {year}
  {2013})}\BibitemShut {NoStop}%
\bibitem [{\citenamefont {Kalb}\ \emph {et~al.}(2017)\citenamefont {Kalb},
  \citenamefont {Reiserer}, \citenamefont {Humphreys}, \citenamefont
  {Bakermans}, \citenamefont {Kamerling}, \citenamefont {Nickerson},
  \citenamefont {Benjamin}, \citenamefont {Twitchen}, \citenamefont {Markham},\
  and\ \citenamefont {Hanson}}]{kalb2017entanglement}%
  \BibitemOpen
  \bibfield  {author} {\bibinfo {author} {\bibfnamefont {N.}~\bibnamefont
  {Kalb}}, \bibinfo {author} {\bibfnamefont {A.~A.}\ \bibnamefont {Reiserer}},
  \bibinfo {author} {\bibfnamefont {P.~C.}\ \bibnamefont {Humphreys}}, \bibinfo
  {author} {\bibfnamefont {J.~J.}\ \bibnamefont {Bakermans}}, \bibinfo {author}
  {\bibfnamefont {S.~J.}\ \bibnamefont {Kamerling}}, \bibinfo {author}
  {\bibfnamefont {N.~H.}\ \bibnamefont {Nickerson}}, \bibinfo {author}
  {\bibfnamefont {S.~C.}\ \bibnamefont {Benjamin}}, \bibinfo {author}
  {\bibfnamefont {D.~J.}\ \bibnamefont {Twitchen}}, \bibinfo {author}
  {\bibfnamefont {M.}~\bibnamefont {Markham}},\ and\ \bibinfo {author}
  {\bibfnamefont {R.}~\bibnamefont {Hanson}},\ }\bibfield  {title} {\bibinfo
  {title} {Entanglement distillation between solid-state quantum network
  nodes},\ }\href@noop {} {\bibfield  {journal} {\bibinfo  {journal} {Science}\
  }\textbf {\bibinfo {volume} {356}},\ \bibinfo {pages} {928} (\bibinfo {year}
  {2017})}\BibitemShut {NoStop}%
\bibitem [{\citenamefont {Waldherr}\ \emph {et~al.}(2014)\citenamefont
  {Waldherr}, \citenamefont {Wang}, \citenamefont {Zaiser}, \citenamefont
  {Jamali}, \citenamefont {Schulte-Herbr{\"u}ggen}, \citenamefont {Abe},
  \citenamefont {Ohshima}, \citenamefont {Isoya}, \citenamefont {Du},
  \citenamefont {Neumann} \emph {et~al.}}]{waldherr2014quantum}%
  \BibitemOpen
  \bibfield  {author} {\bibinfo {author} {\bibfnamefont {G.}~\bibnamefont
  {Waldherr}}, \bibinfo {author} {\bibfnamefont {Y.}~\bibnamefont {Wang}},
  \bibinfo {author} {\bibfnamefont {S.}~\bibnamefont {Zaiser}}, \bibinfo
  {author} {\bibfnamefont {M.}~\bibnamefont {Jamali}}, \bibinfo {author}
  {\bibfnamefont {T.}~\bibnamefont {Schulte-Herbr{\"u}ggen}}, \bibinfo {author}
  {\bibfnamefont {H.}~\bibnamefont {Abe}}, \bibinfo {author} {\bibfnamefont
  {T.}~\bibnamefont {Ohshima}}, \bibinfo {author} {\bibfnamefont
  {J.}~\bibnamefont {Isoya}}, \bibinfo {author} {\bibfnamefont
  {J.}~\bibnamefont {Du}}, \bibinfo {author} {\bibfnamefont {P.}~\bibnamefont
  {Neumann}}, \emph {et~al.},\ }\bibfield  {title} {\bibinfo {title} {Quantum
  error correction in a solid-state hybrid spin register},\ }\href@noop {}
  {\bibfield  {journal} {\bibinfo  {journal} {Nature}\ }\textbf {\bibinfo
  {volume} {506}},\ \bibinfo {pages} {204} (\bibinfo {year}
  {2014})}\BibitemShut {NoStop}%
\bibitem [{\citenamefont {Pompili}\ \emph {et~al.}(2021)\citenamefont
  {Pompili}, \citenamefont {Hermans}, \citenamefont {Baier}, \citenamefont
  {Beukers}, \citenamefont {Humphreys}, \citenamefont {Schouten}, \citenamefont
  {Vermeulen}, \citenamefont {Tiggelman}, \citenamefont {dos Santos~Martins},
  \citenamefont {Dirkse} \emph {et~al.}}]{pompili2021realization}%
  \BibitemOpen
  \bibfield  {author} {\bibinfo {author} {\bibfnamefont {M.}~\bibnamefont
  {Pompili}}, \bibinfo {author} {\bibfnamefont {S.~L.}\ \bibnamefont
  {Hermans}}, \bibinfo {author} {\bibfnamefont {S.}~\bibnamefont {Baier}},
  \bibinfo {author} {\bibfnamefont {H.~K.}\ \bibnamefont {Beukers}}, \bibinfo
  {author} {\bibfnamefont {P.~C.}\ \bibnamefont {Humphreys}}, \bibinfo {author}
  {\bibfnamefont {R.~N.}\ \bibnamefont {Schouten}}, \bibinfo {author}
  {\bibfnamefont {R.~F.}\ \bibnamefont {Vermeulen}}, \bibinfo {author}
  {\bibfnamefont {M.~J.}\ \bibnamefont {Tiggelman}}, \bibinfo {author}
  {\bibfnamefont {L.}~\bibnamefont {dos Santos~Martins}}, \bibinfo {author}
  {\bibfnamefont {B.}~\bibnamefont {Dirkse}}, \emph {et~al.},\ }\bibfield
  {title} {\bibinfo {title} {Realization of a multinode quantum network of
  remote solid-state qubits},\ }\href@noop {} {\bibfield  {journal} {\bibinfo
  {journal} {Science}\ }\textbf {\bibinfo {volume} {372}},\ \bibinfo {pages}
  {259} (\bibinfo {year} {2021})}\BibitemShut {NoStop}%
\bibitem [{\citenamefont {Gross}\ \emph {et~al.}(2017)\citenamefont {Gross},
  \citenamefont {Akhtar}, \citenamefont {Garcia}, \citenamefont
  {Mart{\'\i}nez}, \citenamefont {Chouaieb}, \citenamefont {Garcia},
  \citenamefont {Carr{\'e}t{\'e}ro}, \citenamefont {Barth{\'e}l{\'e}my},
  \citenamefont {Appel}, \citenamefont {Maletinsky} \emph
  {et~al.}}]{gross2017real}%
  \BibitemOpen
  \bibfield  {author} {\bibinfo {author} {\bibfnamefont {I.}~\bibnamefont
  {Gross}}, \bibinfo {author} {\bibfnamefont {W.}~\bibnamefont {Akhtar}},
  \bibinfo {author} {\bibfnamefont {V.}~\bibnamefont {Garcia}}, \bibinfo
  {author} {\bibfnamefont {L.}~\bibnamefont {Mart{\'\i}nez}}, \bibinfo {author}
  {\bibfnamefont {S.}~\bibnamefont {Chouaieb}}, \bibinfo {author}
  {\bibfnamefont {K.}~\bibnamefont {Garcia}}, \bibinfo {author} {\bibfnamefont
  {C.}~\bibnamefont {Carr{\'e}t{\'e}ro}}, \bibinfo {author} {\bibfnamefont
  {A.}~\bibnamefont {Barth{\'e}l{\'e}my}}, \bibinfo {author} {\bibfnamefont
  {P.}~\bibnamefont {Appel}}, \bibinfo {author} {\bibfnamefont
  {P.}~\bibnamefont {Maletinsky}}, \emph {et~al.},\ }\bibfield  {title}
  {\bibinfo {title} {Real-space imaging of non-collinear antiferromagnetic
  order with a single-spin magnetometer},\ }\href@noop {} {\bibfield  {journal}
  {\bibinfo  {journal} {Nature}\ }\textbf {\bibinfo {volume} {549}},\ \bibinfo
  {pages} {252} (\bibinfo {year} {2017})}\BibitemShut {NoStop}%
\bibitem [{\citenamefont {Dolde}\ \emph {et~al.}(2011)\citenamefont {Dolde},
  \citenamefont {Fedder}, \citenamefont {Doherty}, \citenamefont {N{\"o}bauer},
  \citenamefont {Rempp}, \citenamefont {Balasubramanian}, \citenamefont {Wolf},
  \citenamefont {Reinhard}, \citenamefont {Hollenberg}, \citenamefont {Jelezko}
  \emph {et~al.}}]{dolde2011electric}%
  \BibitemOpen
  \bibfield  {author} {\bibinfo {author} {\bibfnamefont {F.}~\bibnamefont
  {Dolde}}, \bibinfo {author} {\bibfnamefont {H.}~\bibnamefont {Fedder}},
  \bibinfo {author} {\bibfnamefont {M.~W.}\ \bibnamefont {Doherty}}, \bibinfo
  {author} {\bibfnamefont {T.}~\bibnamefont {N{\"o}bauer}}, \bibinfo {author}
  {\bibfnamefont {F.}~\bibnamefont {Rempp}}, \bibinfo {author} {\bibfnamefont
  {G.}~\bibnamefont {Balasubramanian}}, \bibinfo {author} {\bibfnamefont
  {T.}~\bibnamefont {Wolf}}, \bibinfo {author} {\bibfnamefont {F.}~\bibnamefont
  {Reinhard}}, \bibinfo {author} {\bibfnamefont {L.~C.}\ \bibnamefont
  {Hollenberg}}, \bibinfo {author} {\bibfnamefont {F.}~\bibnamefont {Jelezko}},
  \emph {et~al.},\ }\bibfield  {title} {\bibinfo {title} {Electric-field
  sensing using single diamond spins},\ }\href@noop {} {\bibfield  {journal}
  {\bibinfo  {journal} {Nature Physics}\ }\textbf {\bibinfo {volume} {7}},\
  \bibinfo {pages} {459} (\bibinfo {year} {2011})}\BibitemShut {NoStop}%
\bibitem [{\citenamefont {Hoese}\ \emph {et~al.}(2021)\citenamefont {Hoese},
  \citenamefont {Koch}, \citenamefont {Bharadwaj}, \citenamefont {Lang},
  \citenamefont {Hadden}, \citenamefont {Yoshizaki}, \citenamefont
  {Giakoumaki}, \citenamefont {Ramponi}, \citenamefont {Jelezko}, \citenamefont
  {Eaton} \emph {et~al.}}]{hoese2021integrated}%
  \BibitemOpen
  \bibfield  {author} {\bibinfo {author} {\bibfnamefont {M.}~\bibnamefont
  {Hoese}}, \bibinfo {author} {\bibfnamefont {M.~K.}\ \bibnamefont {Koch}},
  \bibinfo {author} {\bibfnamefont {V.}~\bibnamefont {Bharadwaj}}, \bibinfo
  {author} {\bibfnamefont {J.}~\bibnamefont {Lang}}, \bibinfo {author}
  {\bibfnamefont {J.~P.}\ \bibnamefont {Hadden}}, \bibinfo {author}
  {\bibfnamefont {R.}~\bibnamefont {Yoshizaki}}, \bibinfo {author}
  {\bibfnamefont {A.~N.}\ \bibnamefont {Giakoumaki}}, \bibinfo {author}
  {\bibfnamefont {R.}~\bibnamefont {Ramponi}}, \bibinfo {author} {\bibfnamefont
  {F.}~\bibnamefont {Jelezko}}, \bibinfo {author} {\bibfnamefont {S.~M.}\
  \bibnamefont {Eaton}}, \emph {et~al.},\ }\bibfield  {title} {\bibinfo {title}
  {Integrated magnetometry platform with stackable waveguide-assisted detection
  channels for sensing arrays},\ }\href@noop {} {\bibfield  {journal} {\bibinfo
   {journal} {Physical Review Applied}\ }\textbf {\bibinfo {volume} {15}},\
  \bibinfo {pages} {054059} (\bibinfo {year} {2021})}\BibitemShut {NoStop}%
\bibitem [{\citenamefont {Bai}\ \emph {et~al.}(2020)\citenamefont {Bai},
  \citenamefont {Huynh}, \citenamefont {Simpson}, \citenamefont {Reineck},
  \citenamefont {Vahid}, \citenamefont {Greentree}, \citenamefont {Foster},
  \citenamefont {Ebendorff-Heidepriem},\ and\ \citenamefont
  {Gibson}}]{bai2020fluorescent}%
  \BibitemOpen
  \bibfield  {author} {\bibinfo {author} {\bibfnamefont {D.}~\bibnamefont
  {Bai}}, \bibinfo {author} {\bibfnamefont {M.~H.}\ \bibnamefont {Huynh}},
  \bibinfo {author} {\bibfnamefont {D.~A.}\ \bibnamefont {Simpson}}, \bibinfo
  {author} {\bibfnamefont {P.}~\bibnamefont {Reineck}}, \bibinfo {author}
  {\bibfnamefont {S.~A.}\ \bibnamefont {Vahid}}, \bibinfo {author}
  {\bibfnamefont {A.~D.}\ \bibnamefont {Greentree}}, \bibinfo {author}
  {\bibfnamefont {S.}~\bibnamefont {Foster}}, \bibinfo {author} {\bibfnamefont
  {H.}~\bibnamefont {Ebendorff-Heidepriem}},\ and\ \bibinfo {author}
  {\bibfnamefont {B.}~\bibnamefont {Gibson}},\ }\bibfield  {title} {\bibinfo
  {title} {Fluorescent diamond microparticle doped glass fiber for magnetic
  field sensing},\ }\href@noop {} {\bibfield  {journal} {\bibinfo  {journal}
  {APL Materials}\ }\textbf {\bibinfo {volume} {8}},\ \bibinfo {pages} {081102}
  (\bibinfo {year} {2020})}\BibitemShut {NoStop}%
\bibitem [{\citenamefont {Moody}\ \emph {et~al.}(2022)\citenamefont {Moody},
  \citenamefont {Sorger}, \citenamefont {Blumenthal}, \citenamefont
  {Juodawlkis}, \citenamefont {Loh}, \citenamefont {Sorace-Agaskar},
  \citenamefont {Jones}, \citenamefont {Balram}, \citenamefont {Matthews},
  \citenamefont {Laing} \emph {et~al.}}]{moody20222022}%
  \BibitemOpen
  \bibfield  {author} {\bibinfo {author} {\bibfnamefont {G.}~\bibnamefont
  {Moody}}, \bibinfo {author} {\bibfnamefont {V.~J.}\ \bibnamefont {Sorger}},
  \bibinfo {author} {\bibfnamefont {D.~J.}\ \bibnamefont {Blumenthal}},
  \bibinfo {author} {\bibfnamefont {P.~W.}\ \bibnamefont {Juodawlkis}},
  \bibinfo {author} {\bibfnamefont {W.}~\bibnamefont {Loh}}, \bibinfo {author}
  {\bibfnamefont {C.}~\bibnamefont {Sorace-Agaskar}}, \bibinfo {author}
  {\bibfnamefont {A.~E.}\ \bibnamefont {Jones}}, \bibinfo {author}
  {\bibfnamefont {K.~C.}\ \bibnamefont {Balram}}, \bibinfo {author}
  {\bibfnamefont {J.~C.}\ \bibnamefont {Matthews}}, \bibinfo {author}
  {\bibfnamefont {A.}~\bibnamefont {Laing}}, \emph {et~al.},\ }\bibfield
  {title} {\bibinfo {title} {2022 roadmap on integrated quantum photonics},\
  }\href@noop {} {\bibfield  {journal} {\bibinfo  {journal} {Journal of
  Physics: Photonics}\ }\textbf {\bibinfo {volume} {4}},\ \bibinfo {pages}
  {012501} (\bibinfo {year} {2022})}\BibitemShut {NoStop}%
\bibitem [{\citenamefont {Awschalom}\ \emph {et~al.}(2018)\citenamefont
  {Awschalom}, \citenamefont {Hanson}, \citenamefont {Wrachtrup},\ and\
  \citenamefont {Zhou}}]{awschalom2018quantum}%
  \BibitemOpen
  \bibfield  {author} {\bibinfo {author} {\bibfnamefont {D.~D.}\ \bibnamefont
  {Awschalom}}, \bibinfo {author} {\bibfnamefont {R.}~\bibnamefont {Hanson}},
  \bibinfo {author} {\bibfnamefont {J.}~\bibnamefont {Wrachtrup}},\ and\
  \bibinfo {author} {\bibfnamefont {B.~B.}\ \bibnamefont {Zhou}},\ }\bibfield
  {title} {\bibinfo {title} {Quantum technologies with optically interfaced
  solid-state spins},\ }\href@noop {} {\bibfield  {journal} {\bibinfo
  {journal} {Nature Photonics}\ }\textbf {\bibinfo {volume} {12}},\ \bibinfo
  {pages} {516} (\bibinfo {year} {2018})}\BibitemShut {NoStop}%
\bibitem [{\citenamefont {Kim}\ \emph {et~al.}(2019)\citenamefont {Kim},
  \citenamefont {Ibrahim}, \citenamefont {Foy}, \citenamefont {Trusheim},
  \citenamefont {Han},\ and\ \citenamefont {Englund}}]{kim2019cmos}%
  \BibitemOpen
  \bibfield  {author} {\bibinfo {author} {\bibfnamefont {D.}~\bibnamefont
  {Kim}}, \bibinfo {author} {\bibfnamefont {M.~I.}\ \bibnamefont {Ibrahim}},
  \bibinfo {author} {\bibfnamefont {C.}~\bibnamefont {Foy}}, \bibinfo {author}
  {\bibfnamefont {M.~E.}\ \bibnamefont {Trusheim}}, \bibinfo {author}
  {\bibfnamefont {R.}~\bibnamefont {Han}},\ and\ \bibinfo {author}
  {\bibfnamefont {D.~R.}\ \bibnamefont {Englund}},\ }\bibfield  {title}
  {\bibinfo {title} {A cmos-integrated quantum sensor based on
  nitrogen--vacancy centres},\ }\href@noop {} {\bibfield  {journal} {\bibinfo
  {journal} {Nature Electronics}\ }\textbf {\bibinfo {volume} {2}},\ \bibinfo
  {pages} {284} (\bibinfo {year} {2019})}\BibitemShut {NoStop}%
\bibitem [{\citenamefont {Zhang}\ \emph {et~al.}(2017)\citenamefont {Zhang},
  \citenamefont {Arai}, \citenamefont {Belthangady}, \citenamefont {Jaskula},\
  and\ \citenamefont {Walsworth}}]{zhang2017selective}%
  \BibitemOpen
  \bibfield  {author} {\bibinfo {author} {\bibfnamefont {H.}~\bibnamefont
  {Zhang}}, \bibinfo {author} {\bibfnamefont {K.}~\bibnamefont {Arai}},
  \bibinfo {author} {\bibfnamefont {C.}~\bibnamefont {Belthangady}}, \bibinfo
  {author} {\bibfnamefont {J.-C.}\ \bibnamefont {Jaskula}},\ and\ \bibinfo
  {author} {\bibfnamefont {R.~L.}\ \bibnamefont {Walsworth}},\ }\bibfield
  {title} {\bibinfo {title} {Selective addressing of solid-state spins at the
  nanoscale via magnetic resonance frequency encoding},\ }\href@noop {}
  {\bibfield  {journal} {\bibinfo  {journal} {npj Quantum Information}\
  }\textbf {\bibinfo {volume} {3}},\ \bibinfo {pages} {31} (\bibinfo {year}
  {2017})}\BibitemShut {NoStop}%
\bibitem [{\citenamefont {Wan}\ \emph {et~al.}(2020)\citenamefont {Wan},
  \citenamefont {Lu}, \citenamefont {Chen}, \citenamefont {Walsh},
  \citenamefont {Trusheim}, \citenamefont {De~Santis}, \citenamefont {Bersin},
  \citenamefont {Harris}, \citenamefont {Mouradian}, \citenamefont {Christen}
  \emph {et~al.}}]{wan2020large}%
  \BibitemOpen
  \bibfield  {author} {\bibinfo {author} {\bibfnamefont {N.~H.}\ \bibnamefont
  {Wan}}, \bibinfo {author} {\bibfnamefont {T.-J.}\ \bibnamefont {Lu}},
  \bibinfo {author} {\bibfnamefont {K.~C.}\ \bibnamefont {Chen}}, \bibinfo
  {author} {\bibfnamefont {M.~P.}\ \bibnamefont {Walsh}}, \bibinfo {author}
  {\bibfnamefont {M.~E.}\ \bibnamefont {Trusheim}}, \bibinfo {author}
  {\bibfnamefont {L.}~\bibnamefont {De~Santis}}, \bibinfo {author}
  {\bibfnamefont {E.~A.}\ \bibnamefont {Bersin}}, \bibinfo {author}
  {\bibfnamefont {I.~B.}\ \bibnamefont {Harris}}, \bibinfo {author}
  {\bibfnamefont {S.~L.}\ \bibnamefont {Mouradian}}, \bibinfo {author}
  {\bibfnamefont {I.~R.}\ \bibnamefont {Christen}}, \emph {et~al.},\ }\bibfield
   {title} {\bibinfo {title} {Large-scale integration of artificial atoms in
  hybrid photonic circuits},\ }\href@noop {} {\bibfield  {journal} {\bibinfo
  {journal} {Nature}\ }\textbf {\bibinfo {volume} {583}},\ \bibinfo {pages}
  {226} (\bibinfo {year} {2020})}\BibitemShut {NoStop}%
\bibitem [{\citenamefont {Chakravarthi}\ \emph {et~al.}(2022)\citenamefont
  {Chakravarthi}, \citenamefont {Yama}, \citenamefont {Abulnaga}, \citenamefont
  {Huang}, \citenamefont {Pederson}, \citenamefont {Hestroffer}, \citenamefont
  {Hatami}, \citenamefont {de~Leon},\ and\ \citenamefont
  {Fu}}]{chakravarthi2022hybrid}%
  \BibitemOpen
  \bibfield  {author} {\bibinfo {author} {\bibfnamefont {S.}~\bibnamefont
  {Chakravarthi}}, \bibinfo {author} {\bibfnamefont {N.~S.}\ \bibnamefont
  {Yama}}, \bibinfo {author} {\bibfnamefont {A.}~\bibnamefont {Abulnaga}},
  \bibinfo {author} {\bibfnamefont {D.}~\bibnamefont {Huang}}, \bibinfo
  {author} {\bibfnamefont {C.}~\bibnamefont {Pederson}}, \bibinfo {author}
  {\bibfnamefont {K.}~\bibnamefont {Hestroffer}}, \bibinfo {author}
  {\bibfnamefont {F.}~\bibnamefont {Hatami}}, \bibinfo {author} {\bibfnamefont
  {N.~P.}\ \bibnamefont {de~Leon}},\ and\ \bibinfo {author} {\bibfnamefont
  {K.-M.~C.}\ \bibnamefont {Fu}},\ }\bibfield  {title} {\bibinfo {title}
  {Hybrid integration of gap photonic crystal cavities with silicon-vacancy
  centers in diamond by stamp-transfer},\ }\href@noop {} {\bibfield  {journal}
  {\bibinfo  {journal} {arXiv preprint arXiv:2212.04670}\ } (\bibinfo {year}
  {2022})}\BibitemShut {NoStop}%
\bibitem [{\citenamefont {Narun}\ \emph {et~al.}(2022)\citenamefont {Narun},
  \citenamefont {Fishman}, \citenamefont {Shulevitz}, \citenamefont {Patel},\
  and\ \citenamefont {Bassett}}]{narun2022efficient}%
  \BibitemOpen
  \bibfield  {author} {\bibinfo {author} {\bibfnamefont {L.~R.}\ \bibnamefont
  {Narun}}, \bibinfo {author} {\bibfnamefont {R.~E.}\ \bibnamefont {Fishman}},
  \bibinfo {author} {\bibfnamefont {H.~J.}\ \bibnamefont {Shulevitz}}, \bibinfo
  {author} {\bibfnamefont {R.~N.}\ \bibnamefont {Patel}},\ and\ \bibinfo
  {author} {\bibfnamefont {L.~C.}\ \bibnamefont {Bassett}},\ }\bibfield
  {title} {\bibinfo {title} {Efficient analysis of photoluminescence images for
  the classification of single-photon emitters},\ }\href@noop {} {\bibfield
  {journal} {\bibinfo  {journal} {ACS Photonics}\ }\textbf {\bibinfo {volume}
  {9}},\ \bibinfo {pages} {3540} (\bibinfo {year} {2022})}\BibitemShut
  {NoStop}%
\bibitem [{\citenamefont {Kudyshev}\ \emph {et~al.}(2020)\citenamefont
  {Kudyshev}, \citenamefont {Bogdanov}, \citenamefont {Isacsson}, \citenamefont
  {Kildishev}, \citenamefont {Boltasseva},\ and\ \citenamefont
  {Shalaev}}]{kudyshev2020rapid}%
  \BibitemOpen
  \bibfield  {author} {\bibinfo {author} {\bibfnamefont {Z.~A.}\ \bibnamefont
  {Kudyshev}}, \bibinfo {author} {\bibfnamefont {S.~I.}\ \bibnamefont
  {Bogdanov}}, \bibinfo {author} {\bibfnamefont {T.}~\bibnamefont {Isacsson}},
  \bibinfo {author} {\bibfnamefont {A.~V.}\ \bibnamefont {Kildishev}}, \bibinfo
  {author} {\bibfnamefont {A.}~\bibnamefont {Boltasseva}},\ and\ \bibinfo
  {author} {\bibfnamefont {V.~M.}\ \bibnamefont {Shalaev}},\ }\bibfield
  {title} {\bibinfo {title} {Rapid classification of quantum sources enabled by
  machine learning},\ }\href@noop {} {\bibfield  {journal} {\bibinfo  {journal}
  {Advanced Quantum Technologies}\ }\textbf {\bibinfo {volume} {3}},\ \bibinfo
  {pages} {2000067} (\bibinfo {year} {2020})}\BibitemShut {NoStop}%
\bibitem [{\citenamefont {Wolters}\ \emph {et~al.}(2010)\citenamefont
  {Wolters}, \citenamefont {Schell}, \citenamefont {Kewes}, \citenamefont
  {N{\"u}sse}, \citenamefont {Schoengen}, \citenamefont {D{\"o}scher},
  \citenamefont {Hannappel}, \citenamefont {L{\"o}chel}, \citenamefont
  {Barth},\ and\ \citenamefont {Benson}}]{wolters2010enhancement}%
  \BibitemOpen
  \bibfield  {author} {\bibinfo {author} {\bibfnamefont {J.}~\bibnamefont
  {Wolters}}, \bibinfo {author} {\bibfnamefont {A.~W.}\ \bibnamefont {Schell}},
  \bibinfo {author} {\bibfnamefont {G.}~\bibnamefont {Kewes}}, \bibinfo
  {author} {\bibfnamefont {N.}~\bibnamefont {N{\"u}sse}}, \bibinfo {author}
  {\bibfnamefont {M.}~\bibnamefont {Schoengen}}, \bibinfo {author}
  {\bibfnamefont {H.}~\bibnamefont {D{\"o}scher}}, \bibinfo {author}
  {\bibfnamefont {T.}~\bibnamefont {Hannappel}}, \bibinfo {author}
  {\bibfnamefont {B.}~\bibnamefont {L{\"o}chel}}, \bibinfo {author}
  {\bibfnamefont {M.}~\bibnamefont {Barth}},\ and\ \bibinfo {author}
  {\bibfnamefont {O.}~\bibnamefont {Benson}},\ }\bibfield  {title} {\bibinfo
  {title} {Enhancement of the zero phonon line emission from a single nitrogen
  vacancy center in a nanodiamond via coupling to a photonic crystal cavity},\
  }\href@noop {} {\bibfield  {journal} {\bibinfo  {journal} {Applied Physics
  Letters}\ }\textbf {\bibinfo {volume} {97}},\ \bibinfo {pages} {141108}
  (\bibinfo {year} {2010})}\BibitemShut {NoStop}%
\bibitem [{\citenamefont {Van~der Sar}\ \emph {et~al.}(2009)\citenamefont
  {Van~der Sar}, \citenamefont {Heeres}, \citenamefont {Dmochowski},
  \citenamefont {De~Lange}, \citenamefont {Robledo}, \citenamefont
  {Oosterkamp},\ and\ \citenamefont {Hanson}}]{van2009nanopositioning}%
  \BibitemOpen
  \bibfield  {author} {\bibinfo {author} {\bibfnamefont {T.}~\bibnamefont
  {Van~der Sar}}, \bibinfo {author} {\bibfnamefont {E.}~\bibnamefont {Heeres}},
  \bibinfo {author} {\bibfnamefont {G.}~\bibnamefont {Dmochowski}}, \bibinfo
  {author} {\bibfnamefont {G.}~\bibnamefont {De~Lange}}, \bibinfo {author}
  {\bibfnamefont {L.}~\bibnamefont {Robledo}}, \bibinfo {author} {\bibfnamefont
  {T.}~\bibnamefont {Oosterkamp}},\ and\ \bibinfo {author} {\bibfnamefont
  {R.}~\bibnamefont {Hanson}},\ }\bibfield  {title} {\bibinfo {title}
  {Nanopositioning of a diamond nanocrystal containing a single
  nitrogen-vacancy defect center},\ }\href@noop {} {\bibfield  {journal}
  {\bibinfo  {journal} {Applied Physics Letters}\ }\textbf {\bibinfo {volume}
  {94}},\ \bibinfo {pages} {173104} (\bibinfo {year} {2009})}\BibitemShut
  {NoStop}%
\bibitem [{\citenamefont {Masood}\ \emph {et~al.}(2020)\citenamefont {Masood},
  \citenamefont {Geuzebroek}, \citenamefont {Dumon}, \citenamefont {van~der
  Vliet}, \citenamefont {Artundo}, \citenamefont {Hoofman},\ and\ \citenamefont
  {Jans}}]{masood2020pix4life}%
  \BibitemOpen
  \bibfield  {author} {\bibinfo {author} {\bibfnamefont {A.}~\bibnamefont
  {Masood}}, \bibinfo {author} {\bibfnamefont {D.}~\bibnamefont {Geuzebroek}},
  \bibinfo {author} {\bibfnamefont {P.}~\bibnamefont {Dumon}}, \bibinfo
  {author} {\bibfnamefont {M.}~\bibnamefont {van~der Vliet}}, \bibinfo {author}
  {\bibfnamefont {I.}~\bibnamefont {Artundo}}, \bibinfo {author} {\bibfnamefont
  {R.}~\bibnamefont {Hoofman}},\ and\ \bibinfo {author} {\bibfnamefont
  {H.}~\bibnamefont {Jans}},\ }\bibfield  {title} {\bibinfo {title}
  {Pix4life’s silicon nitride integrated photonics platforms: present status
  and future outlook},\ }in\ \href@noop {} {\emph {\bibinfo {booktitle}
  {Integrated Photonics Platforms: Fundamental Research, Manufacturing and
  Applications}}},\ Vol.\ \bibinfo {volume} {11364}\ (\bibinfo {organization}
  {SPIE},\ \bibinfo {year} {2020})\ pp.\ \bibinfo {pages} {80--88}\BibitemShut
  {NoStop}%
\bibitem [{\citenamefont {Smith}\ \emph {et~al.}(2020)\citenamefont {Smith},
  \citenamefont {Monroy-Ruz}, \citenamefont {Rarity},\ and\ \citenamefont
  {C.~Balram}}]{smith2020single}%
  \BibitemOpen
  \bibfield  {author} {\bibinfo {author} {\bibfnamefont {J.}~\bibnamefont
  {Smith}}, \bibinfo {author} {\bibfnamefont {J.}~\bibnamefont {Monroy-Ruz}},
  \bibinfo {author} {\bibfnamefont {J.~G.}\ \bibnamefont {Rarity}},\ and\
  \bibinfo {author} {\bibfnamefont {K.}~\bibnamefont {C.~Balram}},\ }\bibfield
  {title} {\bibinfo {title} {Single photon emission and single spin coherence
  of a nitrogen vacancy center encapsulated in silicon nitride},\ }\href@noop
  {} {\bibfield  {journal} {\bibinfo  {journal} {Applied Physics Letters}\
  }\textbf {\bibinfo {volume} {116}},\ \bibinfo {pages} {134001} (\bibinfo
  {year} {2020})}\BibitemShut {NoStop}%
\bibitem [{\citenamefont {Schell}\ \emph {et~al.}(2011)\citenamefont {Schell},
  \citenamefont {Kewes}, \citenamefont {Schr{\"o}der}, \citenamefont {Wolters},
  \citenamefont {Aichele},\ and\ \citenamefont {Benson}}]{schell2011scanning}%
  \BibitemOpen
  \bibfield  {author} {\bibinfo {author} {\bibfnamefont {A.~W.}\ \bibnamefont
  {Schell}}, \bibinfo {author} {\bibfnamefont {G.}~\bibnamefont {Kewes}},
  \bibinfo {author} {\bibfnamefont {T.}~\bibnamefont {Schr{\"o}der}}, \bibinfo
  {author} {\bibfnamefont {J.}~\bibnamefont {Wolters}}, \bibinfo {author}
  {\bibfnamefont {T.}~\bibnamefont {Aichele}},\ and\ \bibinfo {author}
  {\bibfnamefont {O.}~\bibnamefont {Benson}},\ }\bibfield  {title} {\bibinfo
  {title} {A scanning probe-based pick-and-place procedure for assembly of
  integrated quantum optical hybrid devices},\ }\href@noop {} {\bibfield
  {journal} {\bibinfo  {journal} {Review of Scientific Instruments}\ }\textbf
  {\bibinfo {volume} {82}},\ \bibinfo {pages} {073709} (\bibinfo {year}
  {2011})}\BibitemShut {NoStop}%
\bibitem [{\citenamefont {Ampem-Lassen}\ \emph {et~al.}(2009)\citenamefont
  {Ampem-Lassen}, \citenamefont {Simpson}, \citenamefont {Gibson},
  \citenamefont {Trpkovski}, \citenamefont {Hossain}, \citenamefont
  {Huntington}, \citenamefont {Ganesan}, \citenamefont {Hollenberg},\ and\
  \citenamefont {Prawer}}]{ampem2009nano}%
  \BibitemOpen
  \bibfield  {author} {\bibinfo {author} {\bibfnamefont {E.}~\bibnamefont
  {Ampem-Lassen}}, \bibinfo {author} {\bibfnamefont {D.}~\bibnamefont
  {Simpson}}, \bibinfo {author} {\bibfnamefont {B.}~\bibnamefont {Gibson}},
  \bibinfo {author} {\bibfnamefont {S.}~\bibnamefont {Trpkovski}}, \bibinfo
  {author} {\bibfnamefont {F.}~\bibnamefont {Hossain}}, \bibinfo {author}
  {\bibfnamefont {S.}~\bibnamefont {Huntington}}, \bibinfo {author}
  {\bibfnamefont {K.}~\bibnamefont {Ganesan}}, \bibinfo {author} {\bibfnamefont
  {L.}~\bibnamefont {Hollenberg}},\ and\ \bibinfo {author} {\bibfnamefont
  {S.}~\bibnamefont {Prawer}},\ }\bibfield  {title} {\bibinfo {title}
  {Nano-manipulation of diamond-based single photon sources},\ }\href@noop {}
  {\bibfield  {journal} {\bibinfo  {journal} {Optics express}\ }\textbf
  {\bibinfo {volume} {17}},\ \bibinfo {pages} {11287} (\bibinfo {year}
  {2009})}\BibitemShut {NoStop}%
\bibitem [{\citenamefont {Fujiwara}\ \emph {et~al.}(2017)\citenamefont
  {Fujiwara}, \citenamefont {Neitzke}, \citenamefont {Schr{\"o}der},
  \citenamefont {Schell}, \citenamefont {Wolters}, \citenamefont {Zheng},
  \citenamefont {Mouradian}, \citenamefont {Almoktar}, \citenamefont
  {Takeuchi}, \citenamefont {Englund} \emph {et~al.}}]{fujiwara2017fiber}%
  \BibitemOpen
  \bibfield  {author} {\bibinfo {author} {\bibfnamefont {M.}~\bibnamefont
  {Fujiwara}}, \bibinfo {author} {\bibfnamefont {O.}~\bibnamefont {Neitzke}},
  \bibinfo {author} {\bibfnamefont {T.}~\bibnamefont {Schr{\"o}der}}, \bibinfo
  {author} {\bibfnamefont {A.~W.}\ \bibnamefont {Schell}}, \bibinfo {author}
  {\bibfnamefont {J.}~\bibnamefont {Wolters}}, \bibinfo {author} {\bibfnamefont
  {J.}~\bibnamefont {Zheng}}, \bibinfo {author} {\bibfnamefont
  {S.}~\bibnamefont {Mouradian}}, \bibinfo {author} {\bibfnamefont
  {M.}~\bibnamefont {Almoktar}}, \bibinfo {author} {\bibfnamefont
  {S.}~\bibnamefont {Takeuchi}}, \bibinfo {author} {\bibfnamefont
  {D.}~\bibnamefont {Englund}}, \emph {et~al.},\ }\bibfield  {title} {\bibinfo
  {title} {Fiber-coupled diamond micro-waveguides toward an efficient quantum
  interface for spin defect centers},\ }\href@noop {} {\bibfield  {journal}
  {\bibinfo  {journal} {ACS omega}\ }\textbf {\bibinfo {volume} {2}},\ \bibinfo
  {pages} {7194} (\bibinfo {year} {2017})}\BibitemShut {NoStop}%
\bibitem [{\citenamefont {Hadden}\ \emph {et~al.}(2018)\citenamefont {Hadden},
  \citenamefont {Bharadwaj}, \citenamefont {Sotillo}, \citenamefont {Rampini},
  \citenamefont {Osellame}, \citenamefont {Witmer}, \citenamefont {Jayakumar},
  \citenamefont {Fernandez}, \citenamefont {Chiappini}, \citenamefont
  {Armellini} \emph {et~al.}}]{hadden2018integrated}%
  \BibitemOpen
  \bibfield  {author} {\bibinfo {author} {\bibfnamefont {J.}~\bibnamefont
  {Hadden}}, \bibinfo {author} {\bibfnamefont {V.}~\bibnamefont {Bharadwaj}},
  \bibinfo {author} {\bibfnamefont {B.}~\bibnamefont {Sotillo}}, \bibinfo
  {author} {\bibfnamefont {S.}~\bibnamefont {Rampini}}, \bibinfo {author}
  {\bibfnamefont {R.}~\bibnamefont {Osellame}}, \bibinfo {author}
  {\bibfnamefont {J.}~\bibnamefont {Witmer}}, \bibinfo {author} {\bibfnamefont
  {H.}~\bibnamefont {Jayakumar}}, \bibinfo {author} {\bibfnamefont
  {T.}~\bibnamefont {Fernandez}}, \bibinfo {author} {\bibfnamefont
  {A.}~\bibnamefont {Chiappini}}, \bibinfo {author} {\bibfnamefont
  {C.}~\bibnamefont {Armellini}}, \emph {et~al.},\ }\bibfield  {title}
  {\bibinfo {title} {Integrated waveguides and deterministically positioned
  nitrogen vacancy centers in diamond created by femtosecond laser writing},\
  }\href@noop {} {\bibfield  {journal} {\bibinfo  {journal} {Optics letters}\
  }\textbf {\bibinfo {volume} {43}},\ \bibinfo {pages} {3586} (\bibinfo {year}
  {2018})}\BibitemShut {NoStop}%
\bibitem [{\citenamefont {Knowles}\ \emph {et~al.}(2014)\citenamefont
  {Knowles}, \citenamefont {Kara},\ and\ \citenamefont
  {Atat{\"u}re}}]{knowles2014observing}%
  \BibitemOpen
  \bibfield  {author} {\bibinfo {author} {\bibfnamefont {H.~S.}\ \bibnamefont
  {Knowles}}, \bibinfo {author} {\bibfnamefont {D.~M.}\ \bibnamefont {Kara}},\
  and\ \bibinfo {author} {\bibfnamefont {M.}~\bibnamefont {Atat{\"u}re}},\
  }\bibfield  {title} {\bibinfo {title} {Observing bulk diamond spin coherence
  in high-purity nanodiamonds},\ }\href@noop {} {\bibfield  {journal} {\bibinfo
   {journal} {Nature materials}\ }\textbf {\bibinfo {volume} {13}},\ \bibinfo
  {pages} {21} (\bibinfo {year} {2014})}\BibitemShut {NoStop}%
\bibitem [{\citenamefont {Heffernan}\ \emph {et~al.}(2017)\citenamefont
  {Heffernan}, \citenamefont {Greentree},\ and\ \citenamefont
  {Gibson}}]{heffernan2017nanodiamond}%
  \BibitemOpen
  \bibfield  {author} {\bibinfo {author} {\bibfnamefont {A.~H.}\ \bibnamefont
  {Heffernan}}, \bibinfo {author} {\bibfnamefont {A.~D.}\ \bibnamefont
  {Greentree}},\ and\ \bibinfo {author} {\bibfnamefont {B.~C.}\ \bibnamefont
  {Gibson}},\ }\bibfield  {title} {\bibinfo {title} {Nanodiamond arrays on
  glass for quantification and fluorescence characterisation},\ }\href@noop {}
  {\bibfield  {journal} {\bibinfo  {journal} {Scientific Reports}\ }\textbf
  {\bibinfo {volume} {7}},\ \bibinfo {pages} {9252} (\bibinfo {year}
  {2017})}\BibitemShut {NoStop}%
\bibitem [{\citenamefont {Berthel}\ \emph {et~al.}(2015)\citenamefont
  {Berthel}, \citenamefont {Mollet}, \citenamefont {Dantelle}, \citenamefont
  {Gacoin}, \citenamefont {Huant},\ and\ \citenamefont
  {Drezet}}]{berthel2015photophysics}%
  \BibitemOpen
  \bibfield  {author} {\bibinfo {author} {\bibfnamefont {M.}~\bibnamefont
  {Berthel}}, \bibinfo {author} {\bibfnamefont {O.}~\bibnamefont {Mollet}},
  \bibinfo {author} {\bibfnamefont {G.}~\bibnamefont {Dantelle}}, \bibinfo
  {author} {\bibfnamefont {T.}~\bibnamefont {Gacoin}}, \bibinfo {author}
  {\bibfnamefont {S.}~\bibnamefont {Huant}},\ and\ \bibinfo {author}
  {\bibfnamefont {A.}~\bibnamefont {Drezet}},\ }\bibfield  {title} {\bibinfo
  {title} {Photophysics of single nitrogen-vacancy centers in diamond
  nanocrystals},\ }\href@noop {} {\bibfield  {journal} {\bibinfo  {journal}
  {Physical Review B}\ }\textbf {\bibinfo {volume} {91}},\ \bibinfo {pages}
  {035308} (\bibinfo {year} {2015})}\BibitemShut {NoStop}%
\bibitem [{\citenamefont {Smith}\ \emph {et~al.}(2021)\citenamefont {Smith},
  \citenamefont {Clear}, \citenamefont {Balram}, \citenamefont {McCutcheon},\
  and\ \citenamefont {Rarity}}]{smith2021nitrogen}%
  \BibitemOpen
  \bibfield  {author} {\bibinfo {author} {\bibfnamefont {J.~A.}\ \bibnamefont
  {Smith}}, \bibinfo {author} {\bibfnamefont {C.}~\bibnamefont {Clear}},
  \bibinfo {author} {\bibfnamefont {K.~C.}\ \bibnamefont {Balram}}, \bibinfo
  {author} {\bibfnamefont {D.~P.}\ \bibnamefont {McCutcheon}},\ and\ \bibinfo
  {author} {\bibfnamefont {J.~G.}\ \bibnamefont {Rarity}},\ }\bibfield  {title}
  {\bibinfo {title} {Nitrogen-vacancy center coupled to an
  ultrasmall-mode-volume cavity: a high-efficiency source of indistinguishable
  photons at 200 k},\ }\href@noop {} {\bibfield  {journal} {\bibinfo  {journal}
  {Physical Review Applied}\ }\textbf {\bibinfo {volume} {15}},\ \bibinfo
  {pages} {034029} (\bibinfo {year} {2021})}\BibitemShut {NoStop}%
\bibitem [{\citenamefont {Schr{\"o}der}\ \emph {et~al.}(2011)\citenamefont
  {Schr{\"o}der}, \citenamefont {G{\"a}deke}, \citenamefont {Banholzer},\ and\
  \citenamefont {Benson}}]{schroder2011ultrabright}%
  \BibitemOpen
  \bibfield  {author} {\bibinfo {author} {\bibfnamefont {T.}~\bibnamefont
  {Schr{\"o}der}}, \bibinfo {author} {\bibfnamefont {F.}~\bibnamefont
  {G{\"a}deke}}, \bibinfo {author} {\bibfnamefont {M.~J.}\ \bibnamefont
  {Banholzer}},\ and\ \bibinfo {author} {\bibfnamefont {O.}~\bibnamefont
  {Benson}},\ }\bibfield  {title} {\bibinfo {title} {Ultrabright and efficient
  single-photon generation based on nitrogen-vacancy centres in nanodiamonds on
  a solid immersion lens},\ }\href@noop {} {\bibfield  {journal} {\bibinfo
  {journal} {New Journal of Physics}\ }\textbf {\bibinfo {volume} {13}},\
  \bibinfo {pages} {055017} (\bibinfo {year} {2011})}\BibitemShut {NoStop}%
\bibitem [{\citenamefont {Schr{\"o}der}\ \emph {et~al.}(2012)\citenamefont
  {Schr{\"o}der}, \citenamefont {Fujiwara}, \citenamefont {Noda}, \citenamefont
  {Zhao}, \citenamefont {Benson},\ and\ \citenamefont
  {Takeuchi}}]{schroder2012nanodiamond}%
  \BibitemOpen
  \bibfield  {author} {\bibinfo {author} {\bibfnamefont {T.}~\bibnamefont
  {Schr{\"o}der}}, \bibinfo {author} {\bibfnamefont {M.}~\bibnamefont
  {Fujiwara}}, \bibinfo {author} {\bibfnamefont {T.}~\bibnamefont {Noda}},
  \bibinfo {author} {\bibfnamefont {H.-Q.}\ \bibnamefont {Zhao}}, \bibinfo
  {author} {\bibfnamefont {O.}~\bibnamefont {Benson}},\ and\ \bibinfo {author}
  {\bibfnamefont {S.}~\bibnamefont {Takeuchi}},\ }\bibfield  {title} {\bibinfo
  {title} {A nanodiamond-tapered fiber system with high single-mode coupling
  efficiency},\ }\href@noop {} {\bibfield  {journal} {\bibinfo  {journal}
  {Optics express}\ }\textbf {\bibinfo {volume} {20}},\ \bibinfo {pages}
  {10490} (\bibinfo {year} {2012})}\BibitemShut {NoStop}%
\bibitem [{\citenamefont {Patel}\ \emph {et~al.}(2016)\citenamefont {Patel},
  \citenamefont {Schr{\"o}der}, \citenamefont {Wan}, \citenamefont {Li},
  \citenamefont {Mouradian}, \citenamefont {Chen},\ and\ \citenamefont
  {Englund}}]{patel2016efficient}%
  \BibitemOpen
  \bibfield  {author} {\bibinfo {author} {\bibfnamefont {R.~N.}\ \bibnamefont
  {Patel}}, \bibinfo {author} {\bibfnamefont {T.}~\bibnamefont {Schr{\"o}der}},
  \bibinfo {author} {\bibfnamefont {N.}~\bibnamefont {Wan}}, \bibinfo {author}
  {\bibfnamefont {L.}~\bibnamefont {Li}}, \bibinfo {author} {\bibfnamefont
  {S.~L.}\ \bibnamefont {Mouradian}}, \bibinfo {author} {\bibfnamefont {E.~H.}\
  \bibnamefont {Chen}},\ and\ \bibinfo {author} {\bibfnamefont {D.~R.}\
  \bibnamefont {Englund}},\ }\bibfield  {title} {\bibinfo {title} {Efficient
  photon coupling from a diamond nitrogen vacancy center by integration with
  silica fiber},\ }\href@noop {} {\bibfield  {journal} {\bibinfo  {journal}
  {Light: Science \& Applications}\ }\textbf {\bibinfo {volume} {5}},\ \bibinfo
  {pages} {e16032} (\bibinfo {year} {2016})}\BibitemShut {NoStop}%
\bibitem [{\citenamefont {Duan}\ \emph {et~al.}(2019)\citenamefont {Duan},
  \citenamefont {Du}, \citenamefont {Kavatamane}, \citenamefont {Arumugam},
  \citenamefont {Tzeng}, \citenamefont {Chang},\ and\ \citenamefont
  {Balasubramanian}}]{duan2019efficient}%
  \BibitemOpen
  \bibfield  {author} {\bibinfo {author} {\bibfnamefont {D.}~\bibnamefont
  {Duan}}, \bibinfo {author} {\bibfnamefont {G.}~\bibnamefont {Du}}, \bibinfo
  {author} {\bibfnamefont {V.~K.}\ \bibnamefont {Kavatamane}}, \bibinfo
  {author} {\bibfnamefont {S.}~\bibnamefont {Arumugam}}, \bibinfo {author}
  {\bibfnamefont {Y.-K.}\ \bibnamefont {Tzeng}}, \bibinfo {author}
  {\bibfnamefont {H.-C.}\ \bibnamefont {Chang}},\ and\ \bibinfo {author}
  {\bibfnamefont {G.}~\bibnamefont {Balasubramanian}},\ }\bibfield  {title}
  {\bibinfo {title} {Efficient nitrogen-vacancy centers’ fluorescence
  excitation and collection from micrometer-sized diamond by a tapered optical
  fiber in endoscope-type configuration},\ }\href@noop {} {\bibfield  {journal}
  {\bibinfo  {journal} {Optics Express}\ }\textbf {\bibinfo {volume} {27}},\
  \bibinfo {pages} {6734} (\bibinfo {year} {2019})}\BibitemShut {NoStop}%
\bibitem [{\citenamefont {Kurtsiefer}\ \emph {et~al.}(2000)\citenamefont
  {Kurtsiefer}, \citenamefont {Mayer}, \citenamefont {Zarda},\ and\
  \citenamefont {Weinfurter}}]{kurtsiefer2000stable}%
  \BibitemOpen
  \bibfield  {author} {\bibinfo {author} {\bibfnamefont {C.}~\bibnamefont
  {Kurtsiefer}}, \bibinfo {author} {\bibfnamefont {S.}~\bibnamefont {Mayer}},
  \bibinfo {author} {\bibfnamefont {P.}~\bibnamefont {Zarda}},\ and\ \bibinfo
  {author} {\bibfnamefont {H.}~\bibnamefont {Weinfurter}},\ }\bibfield  {title}
  {\bibinfo {title} {Stable solid-state source of single photons},\ }\href@noop
  {} {\bibfield  {journal} {\bibinfo  {journal} {Physical review letters}\
  }\textbf {\bibinfo {volume} {85}},\ \bibinfo {pages} {290} (\bibinfo {year}
  {2000})}\BibitemShut {NoStop}%
\bibitem [{\citenamefont {Inam}\ \emph
  {et~al.}(2013{\natexlab{a}})\citenamefont {Inam}, \citenamefont {Edmonds},
  \citenamefont {Steel},\ and\ \citenamefont {Castelletto}}]{inam2013tracking}%
  \BibitemOpen
  \bibfield  {author} {\bibinfo {author} {\bibfnamefont {F.~A.}\ \bibnamefont
  {Inam}}, \bibinfo {author} {\bibfnamefont {A.~M.}\ \bibnamefont {Edmonds}},
  \bibinfo {author} {\bibfnamefont {M.~J.}\ \bibnamefont {Steel}},\ and\
  \bibinfo {author} {\bibfnamefont {S.}~\bibnamefont {Castelletto}},\
  }\bibfield  {title} {\bibinfo {title} {Tracking emission rate dynamics of
  nitrogen vacancy centers in nanodiamonds},\ }\href@noop {} {\bibfield
  {journal} {\bibinfo  {journal} {Applied Physics Letters}\ }\textbf {\bibinfo
  {volume} {102}},\ \bibinfo {pages} {253109} (\bibinfo {year}
  {2013}{\natexlab{a}})}\BibitemShut {NoStop}%
\bibitem [{\citenamefont {Bradac}\ \emph {et~al.}(2010)\citenamefont {Bradac},
  \citenamefont {Gaebel}, \citenamefont {Naidoo}, \citenamefont {Sellars},
  \citenamefont {Twamley}, \citenamefont {Brown}, \citenamefont {Barnard},
  \citenamefont {Plakhotnik}, \citenamefont {Zvyagin},\ and\ \citenamefont
  {Rabeau}}]{bradac2010observation}%
  \BibitemOpen
  \bibfield  {author} {\bibinfo {author} {\bibfnamefont {C.}~\bibnamefont
  {Bradac}}, \bibinfo {author} {\bibfnamefont {T.}~\bibnamefont {Gaebel}},
  \bibinfo {author} {\bibfnamefont {N.}~\bibnamefont {Naidoo}}, \bibinfo
  {author} {\bibfnamefont {M.}~\bibnamefont {Sellars}}, \bibinfo {author}
  {\bibfnamefont {J.}~\bibnamefont {Twamley}}, \bibinfo {author} {\bibfnamefont
  {L.}~\bibnamefont {Brown}}, \bibinfo {author} {\bibfnamefont
  {A.}~\bibnamefont {Barnard}}, \bibinfo {author} {\bibfnamefont
  {T.}~\bibnamefont {Plakhotnik}}, \bibinfo {author} {\bibfnamefont
  {A.}~\bibnamefont {Zvyagin}},\ and\ \bibinfo {author} {\bibfnamefont
  {J.}~\bibnamefont {Rabeau}},\ }\bibfield  {title} {\bibinfo {title}
  {Observation and control of blinking nitrogen-vacancy centres in discrete
  nanodiamonds},\ }\href@noop {} {\bibfield  {journal} {\bibinfo  {journal}
  {Nature nanotechnology}\ }\textbf {\bibinfo {volume} {5}},\ \bibinfo {pages}
  {345} (\bibinfo {year} {2010})}\BibitemShut {NoStop}%
\bibitem [{\citenamefont {Brouri}\ \emph {et~al.}(2000)\citenamefont {Brouri},
  \citenamefont {Beveratos}, \citenamefont {Poizat},\ and\ \citenamefont
  {Grangier}}]{Brouri2000Photon}%
  \BibitemOpen
  \bibfield  {author} {\bibinfo {author} {\bibfnamefont {R.}~\bibnamefont
  {Brouri}}, \bibinfo {author} {\bibfnamefont {A.}~\bibnamefont {Beveratos}},
  \bibinfo {author} {\bibfnamefont {J.-P.}\ \bibnamefont {Poizat}},\ and\
  \bibinfo {author} {\bibfnamefont {P.}~\bibnamefont {Grangier}},\ }\bibfield
  {title} {\bibinfo {title} {Photon antibunching in the fluorescence of
  individual color centers in diamond},\ }\href
  {https://doi.org/10.1364/OL.25.001294} {\bibfield  {journal} {\bibinfo
  {journal} {Opt. Lett.}\ }\textbf {\bibinfo {volume} {25}},\ \bibinfo {pages}
  {1294} (\bibinfo {year} {2000})}\BibitemShut {NoStop}%
\bibitem [{\citenamefont {Snyder}\ \emph {et~al.}(1983)\citenamefont {Snyder},
  \citenamefont {Love} \emph {et~al.}}]{snyder1983optical}%
  \BibitemOpen
  \bibfield  {author} {\bibinfo {author} {\bibfnamefont {A.~W.}\ \bibnamefont
  {Snyder}}, \bibinfo {author} {\bibfnamefont {J.~D.}\ \bibnamefont {Love}},
  \emph {et~al.},\ }\href@noop {} {\emph {\bibinfo {title} {Optical waveguide
  theory}}},\ Vol.\ \bibinfo {volume} {175}\ (\bibinfo  {publisher} {Chapman
  and hall London},\ \bibinfo {year} {1983})\BibitemShut {NoStop}%
\bibitem [{\citenamefont {Mohtashami}\ and\ \citenamefont
  {Koenderink}(2013)}]{mohtashami2013suitability}%
  \BibitemOpen
  \bibfield  {author} {\bibinfo {author} {\bibfnamefont {A.}~\bibnamefont
  {Mohtashami}}\ and\ \bibinfo {author} {\bibfnamefont {A.~F.}\ \bibnamefont
  {Koenderink}},\ }\bibfield  {title} {\bibinfo {title} {Suitability of
  nanodiamond nitrogen--vacancy centers for spontaneous emission control
  experiments},\ }\href@noop {} {\bibfield  {journal} {\bibinfo  {journal} {New
  Journal of Physics}\ }\textbf {\bibinfo {volume} {15}},\ \bibinfo {pages}
  {043017} (\bibinfo {year} {2013})}\BibitemShut {NoStop}%
\bibitem [{\citenamefont {Plakhotnik}\ and\ \citenamefont
  {Aman}(2018)}]{plakhotnik2018nv}%
  \BibitemOpen
  \bibfield  {author} {\bibinfo {author} {\bibfnamefont {T.}~\bibnamefont
  {Plakhotnik}}\ and\ \bibinfo {author} {\bibfnamefont {H.}~\bibnamefont
  {Aman}},\ }\bibfield  {title} {\bibinfo {title} {{NV}-centers in
  nanodiamonds: How good they are},\ }\href@noop {} {\bibfield  {journal}
  {\bibinfo  {journal} {Diamond and Related Materials}\ }\textbf {\bibinfo
  {volume} {82}},\ \bibinfo {pages} {87} (\bibinfo {year} {2018})}\BibitemShut
  {NoStop}%
\bibitem [{\citenamefont {Inam}\ \emph
  {et~al.}(2013{\natexlab{b}})\citenamefont {Inam}, \citenamefont {Grogan},
  \citenamefont {Rollings}, \citenamefont {Gaebel}, \citenamefont {Say},
  \citenamefont {Bradac}, \citenamefont {Birks}, \citenamefont {Wadsworth},
  \citenamefont {Castelletto}, \citenamefont {Rabeau} \emph
  {et~al.}}]{inam2013emission}%
  \BibitemOpen
  \bibfield  {author} {\bibinfo {author} {\bibfnamefont {F.~A.}\ \bibnamefont
  {Inam}}, \bibinfo {author} {\bibfnamefont {M.~D.}\ \bibnamefont {Grogan}},
  \bibinfo {author} {\bibfnamefont {M.}~\bibnamefont {Rollings}}, \bibinfo
  {author} {\bibfnamefont {T.}~\bibnamefont {Gaebel}}, \bibinfo {author}
  {\bibfnamefont {J.~M.}\ \bibnamefont {Say}}, \bibinfo {author} {\bibfnamefont
  {C.}~\bibnamefont {Bradac}}, \bibinfo {author} {\bibfnamefont {T.~A.}\
  \bibnamefont {Birks}}, \bibinfo {author} {\bibfnamefont {W.~J.}\ \bibnamefont
  {Wadsworth}}, \bibinfo {author} {\bibfnamefont {S.}~\bibnamefont
  {Castelletto}}, \bibinfo {author} {\bibfnamefont {J.~R.}\ \bibnamefont
  {Rabeau}}, \emph {et~al.},\ }\bibfield  {title} {\bibinfo {title} {Emission
  and nonradiative decay of nanodiamond nv centers in a low refractive index
  environment},\ }\href@noop {} {\bibfield  {journal} {\bibinfo  {journal} {ACS
  nano}\ }\textbf {\bibinfo {volume} {7}},\ \bibinfo {pages} {3833} (\bibinfo
  {year} {2013}{\natexlab{b}})}\BibitemShut {NoStop}%
\bibitem [{\citenamefont {Albrecht}\ \emph {et~al.}(2013)\citenamefont
  {Albrecht}, \citenamefont {Bommer}, \citenamefont {Deutsch}, \citenamefont
  {Reichel},\ and\ \citenamefont {Becher}}]{albrecht2013coupling}%
  \BibitemOpen
  \bibfield  {author} {\bibinfo {author} {\bibfnamefont {R.}~\bibnamefont
  {Albrecht}}, \bibinfo {author} {\bibfnamefont {A.}~\bibnamefont {Bommer}},
  \bibinfo {author} {\bibfnamefont {C.}~\bibnamefont {Deutsch}}, \bibinfo
  {author} {\bibfnamefont {J.}~\bibnamefont {Reichel}},\ and\ \bibinfo {author}
  {\bibfnamefont {C.}~\bibnamefont {Becher}},\ }\bibfield  {title} {\bibinfo
  {title} {Coupling of a single nitrogen-vacancy center in diamond to a
  fiber-based microcavity},\ }\href@noop {} {\bibfield  {journal} {\bibinfo
  {journal} {Physical review letters}\ }\textbf {\bibinfo {volume} {110}},\
  \bibinfo {pages} {243602} (\bibinfo {year} {2013})}\BibitemShut {NoStop}%
\bibitem [{\citenamefont {Fehler}\ \emph {et~al.}(2019)\citenamefont {Fehler},
  \citenamefont {Ovvyan}, \citenamefont {Gruhler}, \citenamefont {Pernice},\
  and\ \citenamefont {Kubanek}}]{fehler2019efficient}%
  \BibitemOpen
  \bibfield  {author} {\bibinfo {author} {\bibfnamefont {K.~G.}\ \bibnamefont
  {Fehler}}, \bibinfo {author} {\bibfnamefont {A.~P.}\ \bibnamefont {Ovvyan}},
  \bibinfo {author} {\bibfnamefont {N.}~\bibnamefont {Gruhler}}, \bibinfo
  {author} {\bibfnamefont {W.~H.}\ \bibnamefont {Pernice}},\ and\ \bibinfo
  {author} {\bibfnamefont {A.}~\bibnamefont {Kubanek}},\ }\bibfield  {title}
  {\bibinfo {title} {Efficient coupling of an ensemble of nitrogen vacancy
  center to the mode of a high-q, si3n4 photonic crystal cavity},\ }\href@noop
  {} {\bibfield  {journal} {\bibinfo  {journal} {ACS nano}\ }\textbf {\bibinfo
  {volume} {13}},\ \bibinfo {pages} {6891} (\bibinfo {year}
  {2019})}\BibitemShut {NoStop}%
\bibitem [{\citenamefont {Hiscocks}\ \emph {et~al.}(2009)\citenamefont
  {Hiscocks}, \citenamefont {Su}, \citenamefont {Gibson}, \citenamefont
  {Greentree}, \citenamefont {Hollenberg},\ and\ \citenamefont
  {Ladouceur}}]{Hiscocks2009Slot}%
  \BibitemOpen
  \bibfield  {author} {\bibinfo {author} {\bibfnamefont {M.~P.}\ \bibnamefont
  {Hiscocks}}, \bibinfo {author} {\bibfnamefont {C.-H.}\ \bibnamefont {Su}},
  \bibinfo {author} {\bibfnamefont {B.~C.}\ \bibnamefont {Gibson}}, \bibinfo
  {author} {\bibfnamefont {A.~D.}\ \bibnamefont {Greentree}}, \bibinfo {author}
  {\bibfnamefont {L.~C.~L.}\ \bibnamefont {Hollenberg}},\ and\ \bibinfo
  {author} {\bibfnamefont {F.}~\bibnamefont {Ladouceur}},\ }\bibfield  {title}
  {\bibinfo {title} {Slot-waveguide cavities for optical quantum information
  applications},\ }\href {https://doi.org/10.1364/OE.17.007295} {\bibfield
  {journal} {\bibinfo  {journal} {Opt. Express}\ }\textbf {\bibinfo {volume}
  {17}},\ \bibinfo {pages} {7295} (\bibinfo {year} {2009})}\BibitemShut
  {NoStop}%
\bibitem [{\citenamefont {Wolters}\ \emph {et~al.}(2012)\citenamefont
  {Wolters}, \citenamefont {Kewes}, \citenamefont {Schell}, \citenamefont
  {N{\"u}sse}, \citenamefont {Schoengen}, \citenamefont {L{\"o}chel},
  \citenamefont {Hanke}, \citenamefont {Bratschitsch}, \citenamefont
  {Leitenstorfer}, \citenamefont {Aichele} \emph
  {et~al.}}]{wolters2012coupling}%
  \BibitemOpen
  \bibfield  {author} {\bibinfo {author} {\bibfnamefont {J.}~\bibnamefont
  {Wolters}}, \bibinfo {author} {\bibfnamefont {G.}~\bibnamefont {Kewes}},
  \bibinfo {author} {\bibfnamefont {A.~W.}\ \bibnamefont {Schell}}, \bibinfo
  {author} {\bibfnamefont {N.}~\bibnamefont {N{\"u}sse}}, \bibinfo {author}
  {\bibfnamefont {M.}~\bibnamefont {Schoengen}}, \bibinfo {author}
  {\bibfnamefont {B.}~\bibnamefont {L{\"o}chel}}, \bibinfo {author}
  {\bibfnamefont {T.}~\bibnamefont {Hanke}}, \bibinfo {author} {\bibfnamefont
  {R.}~\bibnamefont {Bratschitsch}}, \bibinfo {author} {\bibfnamefont
  {A.}~\bibnamefont {Leitenstorfer}}, \bibinfo {author} {\bibfnamefont
  {T.}~\bibnamefont {Aichele}}, \emph {et~al.},\ }\href@noop {} {\bibinfo
  {title} {Coupling of single nitrogen-vacancy defect centers in diamond
  nanocrystals to optical antennas and photonic crystal cavities}} (\bibinfo
  {year} {2012})\BibitemShut {NoStop}%
\bibitem [{\citenamefont {Van~der Sar}\ \emph {et~al.}(2011)\citenamefont
  {Van~der Sar}, \citenamefont {Hagemeier}, \citenamefont {Pfaff},
  \citenamefont {Heeres}, \citenamefont {Thon}, \citenamefont {Kim},
  \citenamefont {Petroff}, \citenamefont {Oosterkamp}, \citenamefont
  {Bouwmeester},\ and\ \citenamefont {Hanson}}]{van2011deterministic}%
  \BibitemOpen
  \bibfield  {author} {\bibinfo {author} {\bibfnamefont {T.}~\bibnamefont
  {Van~der Sar}}, \bibinfo {author} {\bibfnamefont {J.}~\bibnamefont
  {Hagemeier}}, \bibinfo {author} {\bibfnamefont {W.}~\bibnamefont {Pfaff}},
  \bibinfo {author} {\bibfnamefont {E.}~\bibnamefont {Heeres}}, \bibinfo
  {author} {\bibfnamefont {S.}~\bibnamefont {Thon}}, \bibinfo {author}
  {\bibfnamefont {H.}~\bibnamefont {Kim}}, \bibinfo {author} {\bibfnamefont
  {P.}~\bibnamefont {Petroff}}, \bibinfo {author} {\bibfnamefont
  {T.}~\bibnamefont {Oosterkamp}}, \bibinfo {author} {\bibfnamefont
  {D.}~\bibnamefont {Bouwmeester}},\ and\ \bibinfo {author} {\bibfnamefont
  {R.}~\bibnamefont {Hanson}},\ }\bibfield  {title} {\bibinfo {title}
  {Deterministic nanoassembly of a coupled quantum emitter--photonic crystal
  cavity system},\ }\href@noop {} {\bibfield  {journal} {\bibinfo  {journal}
  {Applied Physics Letters}\ }\textbf {\bibinfo {volume} {98}},\ \bibinfo
  {pages} {193103} (\bibinfo {year} {2011})}\BibitemShut {NoStop}%
\bibitem [{\citenamefont {Fujii}\ \emph {et~al.}(2011)\citenamefont {Fujii},
  \citenamefont {Taguchi}, \citenamefont {Saiki},\ and\ \citenamefont
  {Nagasaka}}]{fujii2011fusion}%
  \BibitemOpen
  \bibfield  {author} {\bibinfo {author} {\bibfnamefont {T.}~\bibnamefont
  {Fujii}}, \bibinfo {author} {\bibfnamefont {Y.}~\bibnamefont {Taguchi}},
  \bibinfo {author} {\bibfnamefont {T.}~\bibnamefont {Saiki}},\ and\ \bibinfo
  {author} {\bibfnamefont {Y.}~\bibnamefont {Nagasaka}},\ }\bibfield  {title}
  {\bibinfo {title} {A fusion-spliced near-field optical fiber probe using
  photonic crystal fiber for nanoscale thermometry based on
  fluorescence-lifetime measurement of quantum dots},\ }\href@noop {}
  {\bibfield  {journal} {\bibinfo  {journal} {Sensors}\ }\textbf {\bibinfo
  {volume} {11}},\ \bibinfo {pages} {8358} (\bibinfo {year}
  {2011})}\BibitemShut {NoStop}%
\bibitem [{\citenamefont {Lee}\ \emph {et~al.}(2020)\citenamefont {Lee},
  \citenamefont {Leong}, \citenamefont {Kalashnikov}, \citenamefont {Dai},
  \citenamefont {Gandhi},\ and\ \citenamefont {Krivitsky}}]{lee2020integrated}%
  \BibitemOpen
  \bibfield  {author} {\bibinfo {author} {\bibfnamefont {J.}~\bibnamefont
  {Lee}}, \bibinfo {author} {\bibfnamefont {V.}~\bibnamefont {Leong}}, \bibinfo
  {author} {\bibfnamefont {D.}~\bibnamefont {Kalashnikov}}, \bibinfo {author}
  {\bibfnamefont {J.}~\bibnamefont {Dai}}, \bibinfo {author} {\bibfnamefont
  {A.}~\bibnamefont {Gandhi}},\ and\ \bibinfo {author} {\bibfnamefont {L.~A.}\
  \bibnamefont {Krivitsky}},\ }\bibfield  {title} {\bibinfo {title} {Integrated
  single photon emitters},\ }\href@noop {} {\bibfield  {journal} {\bibinfo
  {journal} {AVS Quantum Science}\ }\textbf {\bibinfo {volume} {2}},\ \bibinfo
  {pages} {031701} (\bibinfo {year} {2020})}\BibitemShut {NoStop}%
\bibitem [{\citenamefont {Ji}\ \emph {et~al.}(2023)\citenamefont {Ji},
  \citenamefont {Okawachi}, \citenamefont {Gil-Molina}, \citenamefont
  {Corato-Zanarella}, \citenamefont {Roberts}, \citenamefont {Gaeta},\ and\
  \citenamefont {Lipson}}]{ji2023ultra}%
  \BibitemOpen
  \bibfield  {author} {\bibinfo {author} {\bibfnamefont {X.}~\bibnamefont
  {Ji}}, \bibinfo {author} {\bibfnamefont {Y.}~\bibnamefont {Okawachi}},
  \bibinfo {author} {\bibfnamefont {A.}~\bibnamefont {Gil-Molina}}, \bibinfo
  {author} {\bibfnamefont {M.}~\bibnamefont {Corato-Zanarella}}, \bibinfo
  {author} {\bibfnamefont {S.}~\bibnamefont {Roberts}}, \bibinfo {author}
  {\bibfnamefont {A.~L.}\ \bibnamefont {Gaeta}},\ and\ \bibinfo {author}
  {\bibfnamefont {M.}~\bibnamefont {Lipson}},\ }\bibfield  {title} {\bibinfo
  {title} {Ultra-low-loss silicon nitride photonics based on deposited films
  compatible with foundries},\ }\href@noop {} {\bibfield  {journal} {\bibinfo
  {journal} {Laser \& Photonics Reviews}\ ,\ \bibinfo {pages} {2200544}}
  (\bibinfo {year} {2023})}\BibitemShut {NoStop}%
\bibitem [{\citenamefont {Lee}\ \emph {et~al.}(2022)\citenamefont {Lee},
  \citenamefont {Bersin}, \citenamefont {Dahlberg}, \citenamefont {Wehner},\
  and\ \citenamefont {Englund}}]{lee2022quantum}%
  \BibitemOpen
  \bibfield  {author} {\bibinfo {author} {\bibfnamefont {Y.}~\bibnamefont
  {Lee}}, \bibinfo {author} {\bibfnamefont {E.}~\bibnamefont {Bersin}},
  \bibinfo {author} {\bibfnamefont {A.}~\bibnamefont {Dahlberg}}, \bibinfo
  {author} {\bibfnamefont {S.}~\bibnamefont {Wehner}},\ and\ \bibinfo {author}
  {\bibfnamefont {D.}~\bibnamefont {Englund}},\ }\bibfield  {title} {\bibinfo
  {title} {A quantum router architecture for high-fidelity entanglement flows
  in quantum networks},\ }\href@noop {} {\bibfield  {journal} {\bibinfo
  {journal} {npj Quantum Information}\ }\textbf {\bibinfo {volume} {8}},\
  \bibinfo {pages} {75} (\bibinfo {year} {2022})}\BibitemShut {NoStop}%
\bibitem [{\citenamefont {Koong}\ \emph {et~al.}(2020)\citenamefont {Koong},
  \citenamefont {Ballesteros-Garcia}, \citenamefont {Proux}, \citenamefont
  {Dalacu}, \citenamefont {Poole},\ and\ \citenamefont
  {Gerardot}}]{koong2020multiplexed}%
  \BibitemOpen
  \bibfield  {author} {\bibinfo {author} {\bibfnamefont {Z.-X.}\ \bibnamefont
  {Koong}}, \bibinfo {author} {\bibfnamefont {G.}~\bibnamefont
  {Ballesteros-Garcia}}, \bibinfo {author} {\bibfnamefont {R.}~\bibnamefont
  {Proux}}, \bibinfo {author} {\bibfnamefont {D.}~\bibnamefont {Dalacu}},
  \bibinfo {author} {\bibfnamefont {P.~J.}\ \bibnamefont {Poole}},\ and\
  \bibinfo {author} {\bibfnamefont {B.~D.}\ \bibnamefont {Gerardot}},\
  }\bibfield  {title} {\bibinfo {title} {Multiplexed single photons from
  deterministically positioned nanowire quantum dots},\ }\href@noop {}
  {\bibfield  {journal} {\bibinfo  {journal} {Physical Review Applied}\
  }\textbf {\bibinfo {volume} {14}},\ \bibinfo {pages} {034011} (\bibinfo
  {year} {2020})}\BibitemShut {NoStop}%
\bibitem [{\citenamefont {Sharma}\ \emph {et~al.}(2020)\citenamefont {Sharma},
  \citenamefont {Wang}, \citenamefont {Kaushik}, \citenamefont {Cheng},
  \citenamefont {Kumar}, \citenamefont {Wei},\ and\ \citenamefont
  {Li}}]{sharma2020review}%
  \BibitemOpen
  \bibfield  {author} {\bibinfo {author} {\bibfnamefont {T.}~\bibnamefont
  {Sharma}}, \bibinfo {author} {\bibfnamefont {J.}~\bibnamefont {Wang}},
  \bibinfo {author} {\bibfnamefont {B.~K.}\ \bibnamefont {Kaushik}}, \bibinfo
  {author} {\bibfnamefont {Z.}~\bibnamefont {Cheng}}, \bibinfo {author}
  {\bibfnamefont {R.}~\bibnamefont {Kumar}}, \bibinfo {author} {\bibfnamefont
  {Z.}~\bibnamefont {Wei}},\ and\ \bibinfo {author} {\bibfnamefont
  {X.}~\bibnamefont {Li}},\ }\bibfield  {title} {\bibinfo {title} {Review of
  recent progress on silicon nitride-based photonic integrated circuits},\
  }\href@noop {} {\bibfield  {journal} {\bibinfo  {journal} {Ieee Access}\
  }\textbf {\bibinfo {volume} {8}},\ \bibinfo {pages} {195436} (\bibinfo {year}
  {2020})}\BibitemShut {NoStop}%
\end{thebibliography}%

\makeatletter
\renewcommand{\theequation}{S\arabic{equation}}
\renewcommand{\thefigure}{S\arabic{figure}}
\renewcommand{\thetable}{S\arabic{table}}
\pagebreak
\widetext
\begin{center}
\textbf{\large Supplementary Information}
\end{center}

\setcounter{equation}{0}
\setcounter{figure}{0}
\setcounter{table}{0}
\setcounter{page}{1}

\subsection{Main experimental setup}
For experiments in the main text (Fig.3), NV centres are excited by a pulsed 532 nm laser (PDL 800-B; PicoQUANT) for time resolved PL analysis and a CW 532 nm laser (gem 532; Laser Quantum) in other tests. PL is collected through the two grating couplers in the HBT test or single sided through one grating coupler in other measurements.  Coupled PL from the grating coupler passes through an off-chip notch filter (NF01-532U-25; Semrock) and a long pass filter (BLP01-568R-25; Semrockor or FEL-0550; Thorlabs) to block potential pump leakage before detection by single photon avalanche diodes (SPADs)(SPCM-AQRH-12-FC; PerkinElmer). In the HBT test, coincidence counting is  measured with a time-correlated single photon counting system (PicoHarp 300; PicoQUANT).

\subsection{Confocal microscopy setup}
\begin{figure*}[htbp]
\centering
\includegraphics[width=0.8\linewidth]{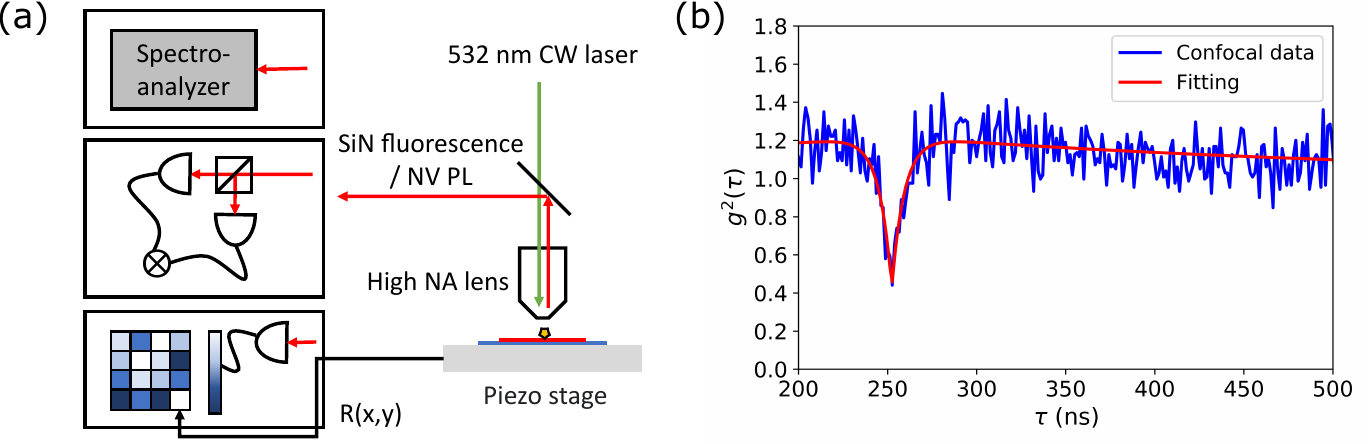}
\caption{Schematic diagram of the confocal microscopy setup and HBT result. (a) The confocal setup used for the spectral measurement, HBT experiment, and imaging. (b) HBT result as measured by the confocal setup.}
\label{fig:conf}
\end{figure*}

A confocal microscope is used to image the PL from NV centres on the chip, to measure the spectrum of SiN fluorescence, and to verify the HBT result. A schematic diagram is shown in Fig.\ref{fig:conf} (a). Using a 0.9 NA microscope objective, the excitation beam is highly focused on the sample producing a nearly diffraction-limited spot ($<$ 1 \textmu m diameter) to excite the NV centre on the sample. The NV centre PL is collected through the same lens and separated from the excitation path by the use of a dichroic mirror. By scanning the position of the sample, a map of detected count rate is generated producing the image of the NV centre at Site A (Fig.2 (c) in the main text). Fluorescence of the SiN film is measured when the SiN waveguide is excited and the spectrum of the fluorescence (Fig.1 (b) in the main text) is analyzed using a spectrometer. A HBT measurement is repeated using the confocal setup by splitting the NV emission upon a beam splitter and analyzing the cross correlation of single photon detection. With the confocal setup, we measure a HBT result in Fig.\ref{fig:conf} (b) with $g^2(0)=0.46$ after background corrections.

\subsection{Noise modeling}
In this section, we model the mechanisms of background noise in our measurements. We show that fluorescence generated in the excitation fibre dominates over the SiN fluorescence in the measured background noise. 

In Fig.3 (b) in the main text, the rate of counts detected at Site B with NDs, compared to the bare waveguide at Site C ($R_\textrm{B}>R_\textrm{C}$), implies that an increased amount of pump or potential fibre fluorescence is coupled into the waveguide when scattered by the NDs. This is compared to -70 dB coupling in the case of a bare waveguide (Fig.1 (c) in the main text). 

In our model, we also estimate the scattering-caused enhancement of coupling. Our estimated noise level agrees with experimental results observed from Fig.3 (b) in the main text.

\begin{figure*}[htbp]
\centering
\includegraphics[width=0.45\linewidth]{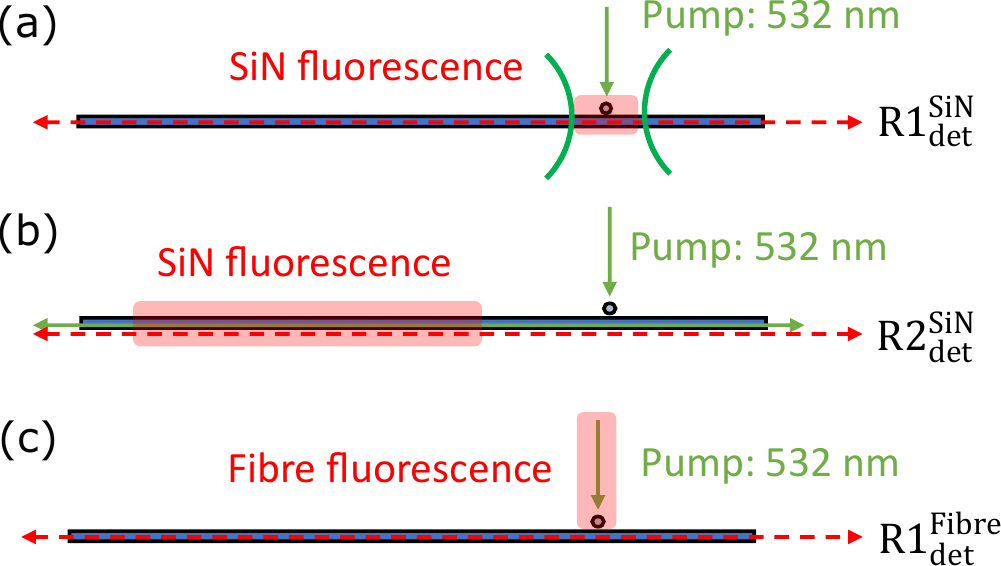}
\caption{Schematic diagram of noise induced by SiN fluorescence and fibre fluorescence. (a) The first term of the SiN fluorescence $R1^{\textrm{SiN}}_{\textrm{det}}$ is where SiN is directly excited by pump (in green) and the SiN fluorescence (in red) is coupled into the waveguide. (b) The second term of the SiN fluorescence $R2^{\textrm{SiN}}_{\textrm{det}}$ corresponds to the small fraction of pump is coupled into the waveguide, inducing SiN fluorescence along the waveguide. (c) The fibre fluorescence term $R1^{\textrm{Fibre}}_{\textrm{det}}$ refers to the fibre fluorescence generated by the pump propagating in the fibre. This fluorescence is also coupled into the waveguide (alongside the pump) through scattering.}
\label{fig:noise_form}
\end{figure*}

The noise contribution from SiN fluorescence is modeled by two terms $R1^{\textrm{SiN}}_{\textrm{det}}$ and $R2^{\textrm{SiN}}_{\textrm{det}}$. The first term (as shown in Fig.\ref{fig:noise_form} (a)) considers the region of SiN waveguide directly excited by the 532 nm pump right at the fibre output. SiN fluorescence is generated in this region and coupled to the waveguide. For the second term (Fig.\ref{fig:noise_form} (b)), a small amount of 532 nm pump is coupled to the waveguide, especially when scattered by NDs. The coupled pump induces SiN fluorescence while propagating along the waveguide. For both terms, the detected count rate for a single grating coupler output (at 6 mW pump) is calculated as:
\begin{equation}
    \begin{split}
    R1^{\textrm{SiN}}_{\textrm{det}}& =R_1^{\textrm{SiN}} \  \eta^{\textrm{SiN}}_{\textrm{wg}} \ \eta^{\textrm{SiN}}_{\textrm{grating}} \ \eta_{\textrm{det}} \\
     & = 2.7 \times 10^{-13} \alpha _1^{\textrm{SiN}} \  \textrm{Hz}, \\
    \end{split}
\label{Eq_R1_Sin}
 \end{equation}
\begin{equation}
    \begin{split}
     R2^{\textrm{SiN}}_{\textrm{det}}&=  R_2^{\textrm{SiN}} \eta^{\textrm{SiN}}_{\textrm{wg}} \ \eta^{\textrm{SiN}}_{\textrm{grating}} \ \eta_{\textrm{det}} \\
     &=2.9 \times 10^{-16} \gamma \alpha_2^{\textrm{SiN}} \  \textrm{Hz}. \\
     \end{split}
 \label{Eq_R2_Sin}
 \end{equation}
where, common to  both equations, $R^{\textrm{SiN}}_{1(2)}$ is the fluorescence count rate (in unit of Hz) generated by excited SiN. $\eta^{\textrm{SiN}}_{\textrm{wg}}$ is the coupling efficiency of SiN fluorescence to the waveguide, split into the $x$ and $-x$ direction and is estimated by simulating dipole emission embedded in the SiN waveguide. $\eta^{\textrm{SiN}}_{\textrm{grating}}$ is the collection efficiency of SiN fluorescence by the grating couplers. The broad spectrum of SiN fluorescence (Fig.1 (b) in the main text) results in a grating coupler efficiency $\eta^{\textrm{SiN}}_{\textrm{grating}}=0.03$. The off-chip detection efficiency remains as $\eta_{\textrm{det}}=0.21$. 

In Eq.\ref{Eq_R1_Sin} and Eq.\ref{Eq_R2_Sin}, the fluorescence count rate can be found as:
 \begin{equation}
     R_{1(2)}^{\textrm{SiN}}=\alpha_{1(2)}^{\textrm{SiN}} E_{\textrm{p1(p2)}} L A,
 \end{equation}
 where  $\alpha_{1(2)}^{\textrm{SiN}}$ (in unit of Hz/W$\cdot$m) is the SiN material response under 532 nm excitation per unit length(depth), $E_{\textrm{p1(p2)}}$ is the power density of excitation beam(in unit of W/m$^2$), $L$ is the penetration depth of pump (in meter) and $A$ is the interaction area (in m$^2$). 

\begin{figure*}[htbp]
\centering
\includegraphics[width=0.55\linewidth]{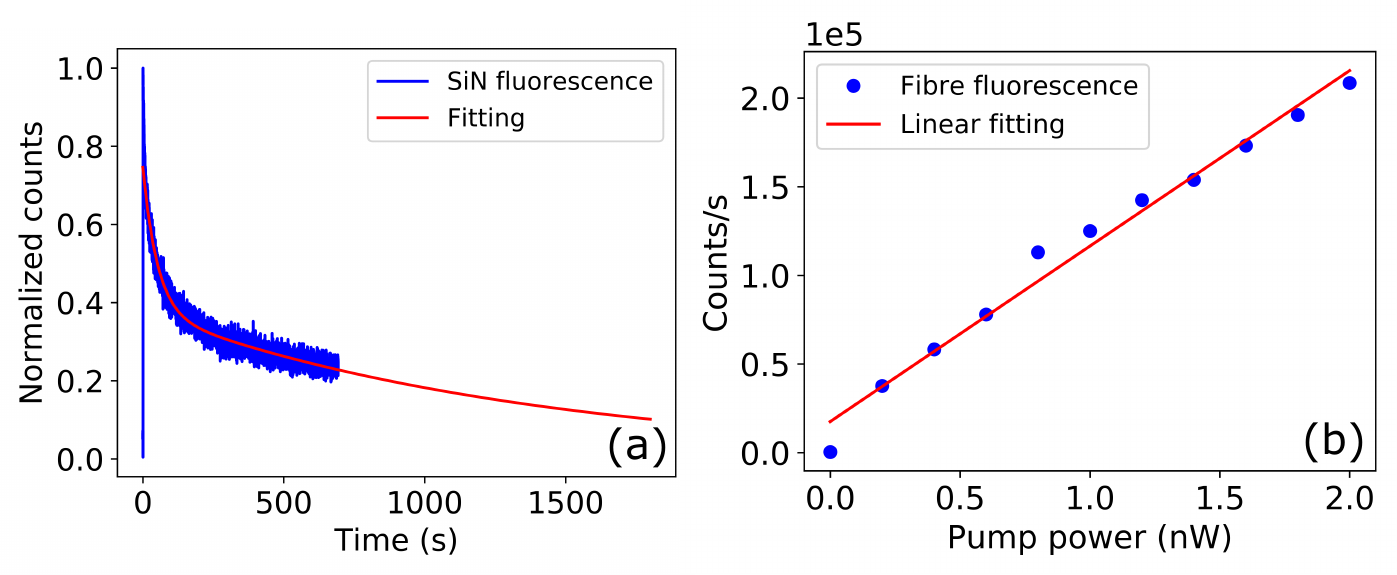}
\caption{Background fluorescence quantification. (a) Bleaching dynamics of SiN fluorescence. The blue curve is measured experimentally and well fitted (in red curve) to a double exponential decay function. (b) Fluorescence from a 5 meter PM630 fibre proportional to the pump power.}
\label{fig:noise_data}
\end{figure*}

As SiN is bleached at the excitation site, this implies $\alpha_1^{\textrm{SiN}} \ll \alpha_2^{\textrm{SiN}}$. In Fig.\ref{fig:noise_data} (a) we measure these bleaching dynamics using a confocal setup when SiN waveguide is excited by 2 mW pump. A long-term exposure to high-power pump results in permanent and strong bleaching of SiN. The value of $\alpha_{1(2)}^{\textrm{SiN}}$ is back calculated from confocal measurement when the SiN waveguide is directly excited. The detected count rate from a bleached region is compared with non-bleached one as:
\begin{equation}
    R_{\textrm{bleached}}=\eta_{\textrm{conf}} \ \alpha_1^{\textrm{SiN}} E_{\textrm{p}} L A,
\label{Eq_conf_bl}
\end{equation}
\begin{equation}
    R_{\textrm{non-bleached}}=\eta_{\textrm{conf}} \ \alpha_2^{\textrm{SiN}} E_{\textrm{p}} L A.
\label{Eq_conf_nbl}
\end{equation}
We record $R_{\textrm{bleached}}=200$ Hz (from Fig.2 (c) in the main text, subtracting 200 Hz detector dark counts) and $R_{\textrm{non-bleached}}=8000$ Hz at 0.6 mW pump. With the setup efficiency $\eta_{\textrm{conf}}=0.01$,
we estimate the SiN response $\alpha_1^{\textrm{SiN}}=1.8 \times 10^{14}$ and $\alpha_2^{\textrm{SiN}}=7 \times 10^{15}$. 

For $R_{2}^{\textrm{SiN}}$, we have $E_{\textrm{p2}}=4 \times 10^{10} \gamma \eta_{\textrm{sc}}$ with the pump-to-waveguide coupling efficiency $\eta_{\textrm{sc}}=10^{-7}/2$ and the scattering-caused enhancement factor $\gamma$. The value of $\gamma$ will vary dependent on the exact make-up of nanodiamonds at each site.
$R1^{\textrm{SiN}}_{\textrm{det}}$ is thus estimated to be 8 Hz and $R2^{\textrm{SiN}}_{\textrm{det}}=0.3 \gamma$ Hz, at 1 mW pump.

As shown in Fig.\ref{fig:noise_form} (c), we expect a strong component of the fluorescence, and hence the measured noise, has been generated externally as the 532 nm laser propagates in the fibre which is then coupled into the SiN waveguide. The detected count rate of fibre fluorescence $R1^{\textrm{Fibre}}_{\textrm{det}}$ is described by
\begin{equation}
    \begin{split}
        R1^{\textrm{Fibre}}_{\textrm{det}}&=R^{\textrm{Fibre}} \ \eta^{\textrm{Fibre}}_{\textrm{wg}} \ \eta^{\textrm{Fibre}}_{\textrm{grating}} \ \eta_{\textrm{det}} \\
         & = 5.2 \times 10^{-10} \gamma R^{\textrm{Fibre}} \eta_{\textrm{det}}  \ \textrm{Hz}.
    \end{split}
     \label{R_fibre}
 \end{equation}
Here, $R^{\textrm{Fibre}}$ is the rate of fibre fluorescence generated. $\eta^{\textrm{Fibre}}_{\textrm{wg}}=\eta_{\textrm{sc}} \gamma$ is the coupling of fibre fluorescence to waveguide, with the same waveguide coupling efficiency $\eta_{\textrm{sc}}=10^{-7}/2$ and scattering enhancement factor $\gamma$ as Eq.\ref{Eq_R2_Sin}. $\eta^{\textrm{Fibre}}_{\textrm{col}}=0.01$ is calculated from the spectrum of fibre fluorescence and $\eta_{\textrm{det}}=0.21$ as before. 

Experimentally, in Fig.\ref{fig:noise_data} 
 (b), we measure the fluorescence of the fibre directly by SPADs at different pump powers using the same pump filter as the main experiment. With the fitted gradient, we extract $R^{\textrm{Fibre}} \eta_{\textrm{det}}= 6 \times 10^{11}$ Hz for a 6 mW pump power. From this, $R1^{\textrm{Fibre}}_{\textrm{det}}=52 \gamma$ Hz is back calculated for a 1 mW pump. 

In conclusion, one can compare the three terms$-$ $R1^{\textrm{SiN}}_{\textrm{det}}=$8 Hz/mW,  $R2^{\textrm{SiN}}_{\textrm{det}}=0.3 \gamma$ Hz/mW, and $R1^{\textrm{Fibre}}_{\textrm{det}}=52 \gamma$ Hz/mW. As the enhancement factor $\gamma$ caused by ND scattering is defined to be greater than one, this implies the fluorescence of fibre dominates in the noise contribution. We estimate the scattering enhancement factor $\gamma=3$, corresponding to a noise level of 60 Hz/mW without NDs ($\gamma=1$) and 165 Hz/mW with NDs ($\gamma=3$). Contribution of each term is shown in Table \ref{Tab: paras} (with $\gamma=3$). One should note that the fibre fluorescence contributes to 95\% (156/(156+8+0.9)) of the noise level. This generally agrees with the experimental observation of 40 Hz/mW (without NDs) and and 160 Hz/mW (with NDs).

\begin{table*}[htbp]
\begin{center}
\renewcommand{\arraystretch}{2}
\begin{tabular}{c|ccc}
 & $R1^{\textrm{SiN}}_{\textrm{det}}$ &  $R2^{\textrm{SiN}}_{\textrm{det}}$ &  $R1^{\textrm{Fibre}}_{\textrm{det}}$ \\ 
 \hline
$R_{1(2)}^{\textrm{SiN}}$ @ 6 mW & $5 \times 10^4$ Hz  & $6.3 \times 10^3$ Hz & - \\
$R^{\textrm{Fibre}}$ @ 6 mW & -  & - & $2.9\times 10^{12}$ Hz \\
$\eta^{\textrm{SiN}}_{\textrm{wg}}$ &   0.15   & 0.15 & -  \\
$\eta^{\textrm{Fibre}}_{\textrm{wg}}$ &  -    & - &  $1.5 \times 10^{-7}$ \\
$\eta^{\textrm{SiN}}_{\textrm{grating}}$ &  0.03    &  0.03 & - \\
$\eta^{\textrm{Fibre}}_{\textrm{grating}}$ &   -   & - &  0.01  \\
$\eta_{\textrm{det}}$ &  0.21    &  0.21 &  0.21 \\
$R_{\textrm{det}}$ @ 1mW &  8 Hz    & 0.9 Hz & 156 Hz \\
\end{tabular}
\renewcommand{\arraystretch}{1}
\caption{\label{Tab: paras} The three terms of noise background $R1^{\textrm{SiN}}_{\textrm{det}}$, $R2^{\textrm{SiN}}_{\textrm{det}}$, and $R1^{\textrm{Fibre}}_{\textrm{det}}$ in comparison (considering $\gamma=3$).}
\end{center}
\end{table*}

\end{document}